\documentclass[aps,pra,twocolumn,groupedaddress,nofootinbib]{revtex4-1}

\usepackage{epsfig,amssymb,amsmath,amsfonts}
\usepackage{times}
\usepackage{graphicx}
\usepackage{epstopdf}

\def\>{\rangle}
\def\<{\langle}

\newcommand{\ignore}[1]{}

\def\bmat{\left[\begin{matrix}} 
\def\emat{\end{matrix}\right]}

\begin{document}

\title{Errors and pseudo-thresholds for incoherent and coherent noise}

\author{Mauricio Guti\'errez$^1$}
\author{Conor Smith$^2$}
\author{Livia Lulushi$^1$}
\author{Smitha Janardan$^1$}
\author{Kenneth R. Brown$^1$}
\email{ken.brown@chemistry.gatech.edu}
\affiliation{$^1$Schools of Chemistry and Biochemistry; Computational Science and Engineering; and Physics\\ 
Georgia Institute of Technology, Atlanta, GA 30332-0400}
\affiliation{$^2$Department of Physics, Lewis \& Clark College, Portland, OR 97219}

\date{\today}

\begin{abstract}

We compare the effect of single qubit incoherent and coherent errors on the logical error rate of the Steane [[7,1,3]] quantum error correction code by performing an exact full-density-matrix simulation of an error correction step. We find that the effective 1-qubit process matrix at the logical level reveals the key differences between the error models and provides insight into why the Pauli twirling approximation is a good approximation for incoherent errors and a poor approximation for coherent ones. Approximate channels composed of Clifford operations and Pauli measurement operators that are pessimistic at the physical level result in pessimistic error rates at the logical level. In addition, we observe that the pseudo-threshold can differ by a factor of five depending on whether the error is calculated using the fidelity or the distance. 

\end{abstract}

\maketitle

\section{Introduction}\label{Sec:Intro}

Error thresholds for fault-tolerant quantum procedures are typically calculated assuming random, independent Pauli errors \cite{Knill, Raussen07, Cross2009, Fowler_threshold}. This error model allows for efficient calculation of thresholds for quantum error correction (QEC) protocols \cite{Gottesmanthesis, CHP, Fast_stabilizer_graph, Aliferis2006, FPT_Smitha}. Actual errors in the laboratory often differ from this approximation and the question remains if these errors are sufficiently small and independent for QEC. There are a number of methods for measuring the error, from the average fidelity \cite{NielsenPLA, Chuang_distance}, which is easy to measure experimentally \cite{Emerson07, RB_Knill2008, ScalableRB_Magesan2011, Magesan_interleavedRB}, to the diamond distance \cite{Chuang_distance, Watrous_semidefinite}, a more challenging measurement. For random Pauli errors, there is no substantial difference between using the average fidelity or the diamond distance to measure the gate error.  Small incoherent errors can be well approximated by Pauli errors.  In turn, these Pauli error models can accurately reproduce the behavior of quantum error-correcting (QEC) protocols in the presence of incoherent channels \cite{Geller_and_Zhou, Cory2014, PRA_us2015, Amara_PTA}.  

On the other hand, if the noise process is best described by a unitary operation, there is a large difference between the error magnitude quantified by the fidelity and the diamond distance \cite{Wallman_boundingerrors2014, Flammia_diamondvsfidelity2015, Sanders_bounding}.   One could generate two distinct Pauli channels to approximate the error.  The first one corresponds to the Pauli twirling approximation, an optimistic model that matches the error in the fidelity.  The second one is a pessimistic model that is constrained to not underestimate a distance-based measure, like the average trace distance or the diamond distance \cite{PRA_us, Cory}.  The latter are referred to as honest approximations.  

The question we address here is how optimistic or pessimistic are these approximations in practice.  We do this in the context of exact simulation of the errors using the Steane [[7,1,3]] code \cite{Steane1996}.  We previously used this code to test expanded efficient error models based on Clifford gates and Pauli measurements for incoherent channels \cite{PRA_us, Cory, PRA_us2015}.  In order to understand the difference between incoherent errors, such as amplitude damping, and coherent errors, such as an unwanted rotation, we start with the respective process matrices.  We then examine the small error limit and perform a polynomial expansion of each entry in the process matrix in terms of the error strength parameter and base our analysis on the leading order terms.  We also construct effective 1-qubit process matrices for a QEC step with the respective error channels acting on the physical qubits.  This method provides a way to visualize how an error channel is modified by QEC.  In real quantum information processing systems, errors arise from both incoherent and coherent processes \cite{Gambetta_IRB_unitary, Laflamme_coherent}.  However, we focus on purely incoherent and coherent errors to emphasize their respective distinctive features.  We include two coherent error models in our analysis, an over-rotation about the $Z$ and the Hadamard, $H$, axes.  Finally, we quantify the error magnitude at the physical and logical levels using various metrics.   

The paper is organized as follows.  In Section \ref{sec:low_error_limit} we introduce our effective 1-qubit process ($\chi$) matrix method and give an illustrative example employing the 3-qubit bit-flip code.  In Section \ref{sec:effective_chi} we present the physical process matrices and the effective 1-qubit process for the model error channels and their approximations in the low noise limit.  We also present the error magnitude of each channel for various error metrics.   In Section \ref{sec:thresholds} we present our level-1 pseudo-threshold values for the Steane [[7,1,3]] code under our model error channels.  We compute the pseudo-threshold based on the different error metrics.  Finally, in Section \ref{sec:conclusions} we conclude and we discuss open questions and future directions.

\section{Analysis in the limit of low error strengths}\label{sec:low_error_limit}






We are interested in understanding why, at the logical error-corrected level, Pauli channels provide exceptionally good approximations to incoherent error channels, but in general a poor one for coherent ones.  Our analysis is based on the effective 1-qubit process matrix for the whole circuit, including the encoding, occurrence of error, syndrome measurement, error correction, and decoding.  This strategy is motivated by the observation that in our simulation scheme the final state of the quantum circuit is always completely localized on the logical codespace, so that the overall circuit can be compactly represented by a 1-qubit process matrix.  

For perfect EC, after the stabilizer measurement and correction, it is evident that the final state will live in the logical codespace.  In general, this will not be the case when the EC is faulty, since errors during the measurement of the stabilizers will cause the logical state to not be projected perfectly onto the code subspaces.  However, in our previous analysis \cite{PRA_us2015}, after faulty EC we always perform one round of perfect EC, to account exclusively for uncorrectable errors.  This has the effect of completely projecting the final state onto the codespace.  Effective noise channels at the logical EC level have been previously employed to study the entanglement in encoded quantum systems \cite{Effective_logical2013} and the error-suppressing properties of the five-qubit code \cite{Fivequbit_Choi}. 

We are particularly interested in the low error rate limit, where it is appropriate to Taylor-expand each entry in the process matrix in terms of powers of the error rate.  This helps us visualize in a very clear way which terms are more important in determining the relevant characteristics of a given error channel.  As an example, consider a 1-qubit coherent error channel consisting of a rotation about the $X$ axis by an angle $\theta$:
\begin{equation}
R_{X}(\theta) = \exp(-i\theta X/2)
\end{equation}

In the normalized Pauli basis $\frac{1}{\sqrt{2}} \lbrace I, X, Y, Z \rbrace$, the process matrix for this channel is:
\begin{equation*}
\setlength{\arraycolsep}{12pt} \begin{pmatrix}
2\cos^2(\theta/2) & i\sin(\theta) & 0 & 0 \\
-i\sin(\theta) & 2\sin^2(\theta/2) & 0 & 0 \\
0 & 0 & 0 & 0 \\
0 & 0 & 0 & 0
\end{pmatrix}
\end{equation*}

In the small error limit ($\theta \rightarrow 0$), this becomes:
\begin{equation*}
\setlength{\arraycolsep}{12pt} \begin{pmatrix}
2 - \theta^2/2 + \mathcal{O}(\theta^4) & i\theta + \mathcal{O}(\theta^3) & 0 & 0 \\
-i\theta + \mathcal{O}(\theta^3) & \theta^2/2 + \mathcal{O}(\theta^4) & 0 & 0 \\
0 & 0 & 0 & 0 \\
0 & 0 & 0 & 0
\end{pmatrix}
\end{equation*}

The Pauli twirled approximation to this channel is given by the diagonal entries of its process matrix:
\begin{equation*}
\setlength{\arraycolsep}{12pt} \begin{pmatrix}
2\cos^2(\theta/2) & 0 & 0 & 0 \\
0 & 2\sin^2(\theta/2) & 0 & 0 \\
0 & 0 & 0 & 0 \\
0 & 0 & 0 & 0
\end{pmatrix}
\end{equation*}

which corresponds to a channel where the qubit is flipped with a probability $p_{x} = \sin^2(\theta/2)$:
\begin{equation} 
\begin{cases}
\sqrt{1-p_{x}} \thinspace I \\
\sqrt{p_{x}} \thinspace X
\end{cases}
\end{equation}

For illustrative purposes, imagine the situation where we use the 3-qubit bit flip code (with stabilizers generated by $ZZI$ and $IZZ$).  We perfectly encode our qubit, then 3 independent error instances happen, 1 on each qubit, and finally we perfectly measure the stabilizer generators, correct, and decode.  If the individual errors correspond to flips with probability $p_{x}$, the effective channel for the whole circuit is given by:
\begin{equation} 
\begin{cases}
\sqrt{(1-p_{x})^3 + 3(1-p_{x})^2 p_{x}} \thinspace I \\
\sqrt{3(1-p_{x})p_{x}^2 + p_{x}^3} \thinspace X
\end{cases}
\end{equation}

This channel is still a probabilistic application of an $X$ operator.  The first Kraus operator corresponds to the situation where either no flip or 1 flip occurred.  The second Kraus operator accounts for the case where 2 or 3 flips occurred, thus causing a logical $X$ error.  If $p_{x} = \sin^2(\theta/2)$, this channel's reduced process matrix is:
\begin{equation*}
\setlength{\arraycolsep}{6pt} \begin{pmatrix}
2(1-p_{x})^{3} + 6(1-p_{x})^{2}p_{x} & 0 \\
0 & 2p_{x}^{3} + 6(1-p_{x})p_{x}^{2}
\end{pmatrix}
\end{equation*}

Here we have only focused on the first 2 rows and columns of the 1-qubit process matrix.  (All the other entries are $0$.)  On the other hand, if the 3 independent errors are coherent rotations about the $X$ axis by an angle $\theta$, the effective reduced process matrix for the whole circuit is:
\begin{equation*}
\setlength{\arraycolsep}{6pt} \begin{pmatrix}
2(1-p_{x})^{3} + 6(1-p_{x})^2 p_{x} & i 8 [p_{x} (1-p_{x})]^{3/2} \\
-i 8  [p_{x} (1-p_{x})]^{3/2}  & 2p_{x}^{3} + 6(1-p_{x}) p_{x}^{2}
\end{pmatrix}
\end{equation*}

Interestingly, for a circuit where the errors are the Pauli twirled approximation to the coherent channels, the effective process matrix for the whole circuit still matches the diagonal entries perfectly.  However, it is completely unable to match the off-diagonal entries.  In the limit of small error, this becomes:
\begin{equation*}
\setlength{\arraycolsep}{6pt} \begin{pmatrix}
2 - (3/8)\theta^4 + \mathcal{O}(\theta^6) & i \theta^3 + \mathcal{O}(\theta^5) \\
-i \theta^3 + \mathcal{O}(\theta^5) & (3/8)\theta^4 + \mathcal{O}(\theta^6)
\end{pmatrix}
\end{equation*}
 
At the physical level, the process matrix for the $R_{X}(\theta)$ channel has diagonal entries proportional to $\theta^2$ and off-diagonal entries proportional to $\theta$.  At the logical level with perfect EC, the leading orders get supressed and the effective process matrix now has diagonal entries proportional to $\theta^4$ and off-diagonal ones proportional to $\theta^3$.  In this case, the Pauli twirled approximation underestimates the magnitude of the error by 1 power of $\theta$ both at the physical and logical levels.    

\section{Effective process matrices for incoherent and coherent channels}\label{sec:effective_chi}

We have followed the same exact full-density-matrix procedure explained in \cite{PRA_us2015} to compute the final states after error correction with the Steane code.  Because of the size of the density matrices and the time it takes to cover all the possible syndrome branches, we are unable to obtain symbolic expressions for the effective 1-qubit process matrices.  Instead, we use quantum process tomography \cite{Pauli_twirling_Chuang, QPTItaly, Lidar_direct, Eugene_direct} to reconstruct the numerical process matrix for various error strengths and subsequently fit each entry to a polynomial to determine the leading order and its coefficient.  We test several polynomial fits and select the one with the smallest total variance.  In every case, the relative variance (variance divided by the value of the leading order coefficient) was less than $10^{-7}$.  Throughout the paper, we will refer to the effective 1-qubit process matrix simply as the process matrix at the logical level.

We report the error magnitude for each channel and its approximations using 3 different metrics: the average error rate (average fidelity error), the average trace distance, and the diamond distance.  Each one of these has a particular importance in our analysis.  The average error rate \cite{Chuang_distance} is the measure of choice when experimentally characterizing quantum gates and it can be efficiently calculated in the laboratory \cite{RB_Knill2008, ScalableRB_Magesan2011}.  Usually in the literature, the error rate is defined with respect to an ideal gate $U$.  Throughout the paper, the ideal gate $U$ will always be the Identity.  There is no loss of generality with this assumption, since it is equivalent to an interaction picture where the error channel $\mathcal{E}$ actually corresponds to the discrepancy channel between the real operation and the ideal gate.   

We define the average error rate, $r$, for a noise process $\mathcal{E}$ as:
\begin{equation}
r(\mathcal{E}) = \langle 1 - F \rangle = 1 - \Bigl \langle \langle \psi | \mathcal{E}(| \psi \rangle \langle \psi |) | \psi \rangle \Bigl \rangle, 
\end{equation}
where the average is defined over the space of pure states\footnote{In our previous work and in older literature, the fidelity between a pure state, $ | \psi \rangle $, and a general state, $\sigma$, was defined as $\sqrt{\langle \psi | \sigma | \psi \rangle}$.  In this paper, we follow the more recent convention and define the fidelity as $\langle \psi | \sigma | \psi \rangle$.}.  Whenever possible, we use the process matrix to analytically integrate over the Bloch sphere surface and obtain the exact expressions for both the average and the standard deviation.  For the cases where the exact symbolic expressions are challenging to compute, we select 150 uniformly distributed states on the surface of the Bloch sphere and integrate numerically.  More specifically, we perform the following procedure: 
\begin{enumerate}
\item For a given initial state, compute the error rate (fidelity error) at 3 different error strengths.
\item Test several polynomial fits, select the one with the smallest total variance, and store the coefficient of the leading order.
\item Repeat this for each one of the 150 initial states.
\item Calculate the average and the standard deviation of the set of 150 leading order coefficients.  
\end{enumerate}   


The average trace distance for the error channel $\mathcal{E}$ is defined as:
\begin{equation}
\langle D_{\textrm{tr}} \rangle = \frac{1}{2} \Bigl \langle \textrm{Tr} | \mathcal{E}(\rho) - \rho | \Bigl \rangle \thickspace , \quad \textrm{for} \quad \rho = | \psi\rangle \langle \psi |.  
\end{equation}
Once again, the average is calculated analytically whenever possible or numerically by following the procedure described above.  The trace distance is important because in our previous work \cite{PRA_us, PRA_us2015} we have used it to distinguish between honest and dishonest approximations to error channels.    
  
Finally, as a worst-case measure with several useful properties, the diamond distance is the preferred metric in the context of fault tolerance \cite{Chuang_distance, Watrous_semidefinite}.  The diamond distance between channels $\mathcal{E}$ and $\mathcal{F}$ is defined as:
\begin{equation}
D_{\diamond} = \frac{1}{2} \underset{\rho}{\textrm{max}} || (\mathcal{E} \otimes I) (\rho) - (\mathcal{F} \otimes I) (\rho) ||_{1}, 
\end{equation}      
where, as mentioned before, the channel $\mathcal{F}$ will always be taken as the Identity.  For a channel acting on a vector space of dimension $d$, the 1-norm is maximized over a vector space of dimension $d^2$.   For a linear operator $A$, the 1-norm is defined as:
\begin{equation}
||A||_{1} = \textrm{Tr} \left [ \sqrt{A^{\dagger} A} \right ]
\end{equation}

To find the coefficient of the leading order term in the diamond distance, we follow the same procedure described previously: we test several polynomial fits and select the one with the smallest total variance. To compute the diamond distance we use QETLAB's diamond norm function \cite{qetlab}.  All the other calculations are done with our own python-based software tools.    

\subsection{Efficiently simulable approximate channels}

In our previous work, we introduced several classically tractable noise models \cite{PRA_us}.   These noise models correspond to Kraus channels where each operator is efficiently simulable in the stabilizer formalism.  The free parameters in our noise models are the probabilities associated with the Kraus operators.  To approximate a target channel, we minimize the Hilbert-Schmidt distance \cite{Distancemeasures} between the process matrices of the classically tractable noise model and the target channel.  For each target non-Clifford error channel, we study two different models: (a) the Pauli channels (PC), which employ only single-qubit Pauli operators, and (b) the expanded channels or Clifford+measurements channels (CMC), which include all the single-qubit Clifford operators and the measurement-induced translations.  When minimizing the Hilbert-Schmidt distance between channels, we have the option to perform a constrained minimization, in which we enforce that for every initial pure state its trace distance to the resulting state after the target transformation is not greater than its trace distance to the resulting state after the model transformation.  Approximate channels that satisfy this condition are referred to as honest, since they do not underestimate the magnitude of the target error.  We label unconstrained approximations as ``a'' and constrained ones as ``w''.  Finally, we also include in our analysis the depolarizing channel (DC), a Pauli channel where the probabilities of the $X$, $Y$, and $Z$ operators are equal.  The approximations are summarized in Table \ref{table:channels_summary}.
\begin{table}[htdp]
\caption{Summary of the various target and approximate channels.}
\begin{center}
\begin{tabular}{| c || c | c|}
\hline
\multicolumn{1}{| c ||}{Channel} & \multicolumn{1}{| c |}{Complete name} & \multicolumn{1}{| c |}{Honesty constrained} \\ \cline{1-3}
ADC & amplitude damping & -- \\ \hline
PolC & polarization along non-Clifford axis & -- \\ \hline
RZC & rotation about the $Z$ axis & -- \\ \hline
RHC & rotation about the $H$ axis & -- \\ \hline
PCa & Pauli & no  \\ \hline
PCw & Pauli & yes  \\ \hline
CMCa & Clifford+measurements & no  \\ \hline
CMCw & Clifford+measurements & yes \\ \hline
DC & Depolarizing channel & no \\ \hline
\end{tabular}
\end{center} \label{table:channels_summary}
\end{table}       

\subsection{Incoherent channels}

We define an incoherent channel as a quantum operation $\mathcal{E}$ that maps, at least, 1 pure state $\rho = | \psi \rangle \langle \psi |$ to a mixed state $\sigma$. As in our previous work \cite{PRA_us, PRA_us2015}, we have selected 2 representative 1-qubit incoherent channels: the amplitude damping channel (ADC) and a depolarizing channel about the non-Clifford $\pi/8$ axis on the $XY$ plane of the Bloch sphere (Pol$_{\pi/8}$C).

\subsubsection{ADC}

\begin{table*}[htdp]
\caption[ADC and approximations]
  {Process matrices for the ADC and its approximations at the physical level and logical level with faulty EC in the low damping limit ($\gamma \rightarrow 0$).  Only the leading orders are shown.}
\begin{center}
\begin{tabular}{ c c c}
\hline \hline
\multicolumn{1}{ c }{Channel} & \multicolumn{1}{ c }{Physical process matrix} & \multicolumn{1}{ c }{Effective process matrix at the logical level} \\ \hline
\\
ADC & $\setlength{\arraycolsep}{13pt} \begin{pmatrix} 2 - \mathcal{O}(\gamma)& 0 & 0 & \gamma/2 \\ 0 & \gamma/2 & -i\gamma/2 & 0 \\ 0 & i\gamma/2 & \gamma/2 & 0 \\ \gamma/2 & 0 & 0 & \gamma^2/8 \end{pmatrix}$ & $\setlength{\arraycolsep}{10.5pt} \begin{pmatrix} 2 - \mathcal{O}(\gamma^2) & 0 & 0 & -211\gamma^3 \\ 0 & 1580\gamma^2 & i211\gamma^3 & 0 \\ 0 & -i211\gamma^3 & 180\gamma^2 & 0 \\ -211\gamma^3 & 0 & 0 & 509\gamma^2 \end{pmatrix}$ \\
\\
PCa & $\setlength{\arraycolsep}{15.25pt} \begin{pmatrix} 2 - \mathcal{O}(\gamma) & 0 & 0 & 0 \\ 0 & \gamma/2 & 0 & 0 \\ 0 & 0 & \gamma/2 & 0 \\ 0 & 0 & 0 & \gamma^2/8 \end{pmatrix}$ &  $\setlength{\arraycolsep}{14.25pt} \begin{pmatrix} 2 - \mathcal{O}(\gamma^2)   & 0 & 0 & 0 \\ 0 & 1570\gamma^2 & 0 & 0 \\ 0 & 0 & 180\gamma^2 & 0 \\ 0 & 0 & 0 & 491\gamma^2 \end{pmatrix}$  \\
\\
PCw & $\setlength{\arraycolsep}{9.25pt} \begin{pmatrix} 2 - \mathcal{O}(\gamma) & 0 & 0 & 0 \\ 0 & 1.047\gamma & 0 & 0 \\ 0 & 0 & 1.047\gamma & 0 \\ 0 & 0 & 0 & 0.2915\gamma \end{pmatrix}$ & $\setlength{\arraycolsep}{14.5pt} \begin{pmatrix} 2 - \mathcal{O}(\gamma^2)  & 0 & 0 & 0 \\ 0 & 7080\gamma^2 & 0 & 0 \\ 0 & 0 & 790\gamma^2 & 0 \\ 0 & 0 & 0 & 3020\gamma^2 \end{pmatrix}$  \\
\\
CMCa & $\setlength{\arraycolsep}{10.5pt} \begin{pmatrix} 2 - \mathcal{O}(\gamma) & 0 & 0 & 3\gamma/8 \\ 0 & 3\gamma/8 & -i3\gamma/8 & 0 \\ 0 & i3\gamma/8 & 3\gamma/8 & 0 \\ 3\gamma/8 & 0 & 0 & 3\gamma/8 \end{pmatrix}$ &  $\setlength{\arraycolsep}{10pt} \begin{pmatrix}  2 - \mathcal{O}(\gamma^2) & 0 & 0 & -89.2\gamma^3 \\ 0 & 988\gamma^2 & i89.2\gamma^3  & 0 \\ 0 & -i89.2\gamma^3 & 101\gamma^2 & 0 \\ -89.2\gamma^3 & 0 & 0 & 771\gamma^2 \end{pmatrix}$  \\
\\
CMCw & $\setlength{\arraycolsep}{13.5pt} \begin{pmatrix} 2 - \mathcal{O}(\gamma) & 0 & 0 & \gamma/2 \\ 0 & \gamma/2 & -i\gamma/2 & 0 \\ 0 & i\gamma/2 & \gamma/2 &0 \\ \gamma/2 & 0 & 0 & \gamma/2 \end{pmatrix}$ & $\setlength{\arraycolsep}{10.5pt} \begin{pmatrix}  2 - \mathcal{O}(\gamma^2) & 0 & 0 & -212\gamma^3 \\ 0 & 1760\gamma^2 & i212\gamma^3 & 0 \\ 0 & -i212\gamma^3 & 180\gamma^2  & 0 \\ -212\gamma^3 & 0 & 0 & 1370\gamma^2 \end{pmatrix}$  \\
\\
DC & $\setlength{\arraycolsep}{16pt} \begin{pmatrix} 2 - \mathcal{O}(\gamma) & 0 & 0 & 0 \\ 0 & \gamma/3 & 0 & 0 \\ 0 & 0 & \gamma/3 & 0 \\ 0 & 0 & 0 & \gamma/3 \end{pmatrix}$ &  $\setlength{\arraycolsep}{14pt} \begin{pmatrix} 2 - \mathcal{O}(\gamma^2)  & 0 & 0 & 0 \\ 0 & 773\gamma^2 & 0 & 0 \\ 0 & 0 & 80.0\gamma^2 & 0 \\ 0 & 0 & 0 & 601\gamma^2 \end{pmatrix}$ 
\\ \\ \hline \hline
\end{tabular}
\end{center} \label{table:process_matrix_ADC}
\end{table*}

\begin{table*}[htdp]
\caption[Behavior of the ADC and its approximations]
{Behavior of the ADC and its approximations at various levels in low damping limit ($\gamma \rightarrow 0$) for the 3 different error metrics.  Only the leading orders are shown.  Standard deviations smaller than $10^{-7}$ are not presented.}
\begin{center}
\begin{tabular}{| c || c | c | c | c | c | c |}
\hline
\multicolumn{1}{| c ||}{Channel} & \multicolumn{3}{| c |}{Physical level} & \multicolumn{3}{| c |}{Logical level with faulty EC} \\ \cline{2-7}
 & $\langle 1 - F \rangle $ & $ \langle D_{\textrm{tr}} \rangle$ & D$_{\diamond}$ & $\langle 1 - F \rangle / 10^3 $ & $\langle D_{\textrm{tr}} \rangle / 10^3 $ & D$_{\diamond} / 10^3$ \\ \hline
ADC & $\left ( 0.33 \pm 0.30 \right ) \gamma$ & $ \left ( 0.55 \pm 0.27 \right ) \gamma$ & $\gamma$ & $ \left ( 0.76 \pm 0.19 \right ) \gamma^2$ & $(0.80 \pm 0.17)\gamma^2$ & $1.14\gamma^2$ \\ \hline
PCa & $\left (0.333 \pm 0.075 \right ) \gamma$ & $ \left ( 0.345 \pm 0.077 \right ) \gamma$ & $\gamma / 2$ & $ \left (0.75  \pm 0.19 \right ) \gamma^2$ & $(0.78 \pm 0.17)\gamma^2$ & $1.12\gamma^2$ \\ \hline
PCw & $ \left (0.80 \pm 0.11 \right ) \gamma$ & $ \left ( 0.81 \pm 0.12 \right ) \gamma$ & $1.19\gamma$ & $  \left ( 3.63 \pm 0.82 \right )  \gamma^2$ & $(3.78 \pm 0.78) \gamma^2$ & $5.45\gamma^2$ \\ \hline
CMCa & $ \left ( 0.38 \pm 0.22 \right ) \gamma$ & $ \left ( 0.50 \pm 0.18 \right ) \gamma$ & $ 3\gamma/4$ & $  \left (0.62  \pm 0.12 \right )  \gamma^2$ & $( 0.64 \pm 0.12) \gamma^2$ & $0.930\gamma^2$ \\ \hline
CMCw & $ \left ( 0.50 \pm 0.29 \right ) \gamma$ & $ \left ( 0.67 \pm 0.24  \right ) \gamma$ & $\gamma$ & $  \left ( 1.10 \pm 0.21 \right ) \gamma^2$ & $( 1.13 \pm 0.22) \gamma^2$ & $1.65\gamma^2$ \\ \hline
DC & $\gamma/3$ & $\gamma/3$ & $\gamma/2$ & $  \left ( 0.485 \pm 0.093 \right )  \gamma^2$ & $(0.497 \pm 0.095) \gamma^2$ & $0.727\gamma^2$ \\ \hline
\end{tabular}
\end{center} \label{table:error_magnitude_ADC}
\end{table*}

Table \ref{table:process_matrix_ADC} presents the process matrices for the ADC and its approximations at the physical level and logical level with faulty EC.  Table \ref{table:error_magnitude_ADC} describes how the error magnitude for these channels scales with the damping strength, $\gamma$, for the metrics introduced in the previous section.  In both tables, the results refer to the behavior in the limit of small damping strength.  For the average error rate and the average trace distance, the standard deviation is also presented.  Standard deviations below $10^{-7}$ are not reported.         

There are several interesting trends.  At the physical level, the entries of the ADC process matrix are all linear in $\gamma$, except for the $\chi_{zz}$ term, which is quadratic.  Consequently, the error magnitude is linear, regardless of the metric used.  At the logical level, all the linear terms are suppressed, which confirms that the Steane code's correcting procedure is indeed fault tolerant and successful in suppressing single errors.  The terms on the diagonal entries are now proportional to $\gamma^2$, while the off-diagonal terms are proportional to $\gamma^3$.  At this level, the error magnitude is quadratic.  

At the physical level, the PCa matches the diagonal entries perfectly, since it corresponds to the Pauli twirling approximation (PTA).  Its average error rate is also identical to that of the ADC, a known property of the PTA.  The average trace distance, however, is less than that of the ADC, and therefore we classify this channel as dishonest.   At the logical level, the PCa still approximates the diagonal terms very closely, but not exactly, which shows that the off-diagonal terms on the physical process matrix can influence the diagonal terms on the logical process matrix.  The error magnitude is practically equivalent to that of the ADC up to second order.  In the case of the PCw, the constrained Pauli approximation, at the physical level all the terms are linear, which guarantees that the error is not underestimated.  At the logical level, this pessimistic behavior becomes even more pronounced, resulting in a very honest, but inaccurate approximation.  

In contrast to the Pauli channels, the expanded (CMC) channels have access to the off-diagonal entries in the process matrix.  At the physical level, this allows both the constrained and unconstrained approximations of the expanded channels to be more accurate.  While an advantage at the physical level, the access to the off-diagonal entries becomes unfavorable at the logical level.  The CMCw is a great example: at the logical level, it matches the off-diagonal terms almost perfectly, but this is useless because its approximation to the leading order in the target channel ($\gamma^2$) is less accurate than that achieved by the PCa, especially on the $\chi_{zz}$ entry.  In other words, for the ADC at the logical level, the critical requirement for a good approximation is to be able to match the diagonal entries accurately, since these contain the leading orders in the error magnitude.     

The constrained approximations at the physical level remain honest at the logical level, for every error measure used.  On the other hand, the unconstrained approximations remain dishonest, with the notable exception of the PCa, which is practically honest up to second order in $\gamma$.  Finally, for every channel and at every level, the average trace distance is consistently about twice as the average error rate.  The diamond distance is about 3-4 times larger than the average error rate.

\subsubsection{Pol$_{\pi/8}$C}

\begin{table*}[htdp]
\caption[PolXY and approximations]
  {Process matrices for the Pol$_{\pi/8}$C and its approximations at the physical level and logical level with faulty EC in the low noise limit ($p \rightarrow 0$).  Only the leading orders are shown.}
\begin{center}
\begin{tabular}{ c c c}
\hline \hline
\multicolumn{1}{ c }{Channel} & \multicolumn{1}{ c }{Physical process matrix} & \multicolumn{1}{ c }{Effective process matrix at the logical level} \\ \hline
\\
Pol$_{\pi/8}$C & $\setlength{\arraycolsep}{14pt} \begin{pmatrix} 2 - \mathcal{O}(p)& 0 & 0 & 0 \\ 0 & (1 + 1/\sqrt{2}) p & (1/\sqrt{2})p & 0 \\ 0 & (1/\sqrt{2})p & (1 - 1/\sqrt{2}) p & 0 \\ 0 & 0 & 0 & 0 \end{pmatrix}$ & $\setlength{\arraycolsep}{12.25pt} \begin{pmatrix} 2 - \mathcal{O}(p^2) & 0 & 0 & -i3790 \, p^4 \\ 0 & 5860 \, p^2 & 337 \, p^3 & 0 \\ 0 & 337 \, p^3 & 61.8 \, p^2 & 0 \\ i3790 \, p^4 & 0 & 0 & 851 \, p^2 \end{pmatrix}$  \\
\\
PCa & $\setlength{\arraycolsep}{14.25pt} \begin{pmatrix} 2 - \mathcal{O}(p) & 0 & 0 & 0 \\ 0 & (1+1/\sqrt{2})p & 0 & 0 \\ 0 & 0 & (1-1/\sqrt{2})p & 0 \\ 0 & 0 & 0 & 0 \end{pmatrix}$ &  $\setlength{\arraycolsep}{14.75pt} \begin{pmatrix} 2 - \mathcal{O}(p^2) & 0 & 0 & 0 \\ 0 & 5860 \, p^2 & 0 & 0 \\ 0 & 0 & 61.8 \, p^2 & 0 \\ 0 & 0 & 0 & 851 \, p^2 \end{pmatrix}$ \\
\\
PCw & $\setlength{\arraycolsep}{9.25pt} \begin{pmatrix} 2 - \mathcal{O}(p) & 0 & 0 & 0 \\ 0 & (1+1/\sqrt{2})p & 0 & 0 \\ 0 & 0 & (1-1/\sqrt{2})p & 0 \\ 0 & 0 & 0 & (1/\sqrt{2})p \end{pmatrix}$ &  $\setlength{\arraycolsep}{13.75pt} \begin{pmatrix} 2 - \mathcal{O}(p^2)  & 0 & 0 & 0 \\ 0 & 6350 \, p^2 & 0 & 0 \\ 0 & 0 & 61.8 \, p^2 & 0 \\ 0 & 0 & 0 & 2190 \, p^2 \end{pmatrix}$ \\
\\
CMCa & $\setlength{\arraycolsep}{10.25pt} \begin{pmatrix} 2 - \mathcal{O}(p) & 0 & 0 & 0 \\ 0 & 3(3 + 1/\sqrt{2})p/7 & (3+1/\sqrt{2})p/7 & 0 \\ 0 & (3+1/\sqrt{2})p/7  & (3+1/\sqrt{2})p/7 & 0 \\ 0 & 0 & 0 & 0  \end{pmatrix}$ &  $\setlength{\arraycolsep}{12.5pt} \begin{pmatrix}  2 - \mathcal{O}(p^2) & 0 & 0 & -i1020 \, p^4  \\ 0 & 6790 \, p^2 & 142 \, p^3 & 0 \\ 0 & 142 \, p^3 & 202 \, p^2 & 0 \\  i1020 \, p^4 & 0 & 0 & 1280 \, p^2 \end{pmatrix}$ \\
\\
CMCw & $\setlength{\arraycolsep}{3pt} \begin{pmatrix} 2 - \mathcal{O}(p) & 0 & 0 & 0 \\ 0 & (1+1/\sqrt{2})p & (1+1/\sqrt{2})p/3 & 0 \\ 0 & (1+1/\sqrt{2})p/3 & (1+1/\sqrt{2})p/3 & 0 \\ 0 & 0 & 0 & (3-2\sqrt{2})p/3 \end{pmatrix}$ &  $\setlength{\arraycolsep}{12.5pt} \begin{pmatrix}  2 - \mathcal{O}(p^2) & 0 & 0 & -i1370 \, p^4 \\ 0 & 7890 \, p^2 & 176 \, p^3 & 0 \\ 0 & 176 \, p^3 & 233 \, p^2 & 0 \\  i1370 \, p^4  & 0 & 0 & 1600 \, p^2 \end{pmatrix}$ \\
\\
DC & $\setlength{\arraycolsep}{22.5pt} \begin{pmatrix} 2 - \mathcal{O}(p) & 0 &0 & 0 \\ 0 & 2p/3 & 0 & 0 \\ 0 & 0 & 2p/3 & 0 \\ 0 & 0 & 0 & 2p/3 \end{pmatrix}$ &  $\setlength{\arraycolsep}{14pt} \begin{pmatrix} 2 - \mathcal{O}(p^2)  & 0 & 0 & 0 \\ 0 & 3090 \, p^2  & 0 & 0 \\ 0 & 0 & 320 \, p^2  & 0 \\ 0 & 0 & 0 & 2400 \, p^2  \end{pmatrix}$ 
\\ \\ \hline \hline
\end{tabular}
\end{center} \label{table:process_matrix_PolXY}
\end{table*}

\begin{table*}[htdp]
\caption[Behavior of the PolC and its approximations at various levels]
{Behavior of the Pol$_{\pi/8}$C and its approximations at various levels in low noise limit ($p \rightarrow 0$) for the 3 different error metrics.  Only the leading orders are shown.  Standard deviations smaller than $10^{-7}$ are not presented.}
\begin{center}
\begin{tabular}{| c || c | c | c | c | c | c |}
\hline
\multicolumn{1}{| c ||}{Channel} & \multicolumn{3}{| c |}{Physical level} & \multicolumn{3}{| c |}{Logical level with faulty EC} \\ \cline{2-7}
 & $\langle 1 - F \rangle$ & $\langle D_{\textrm{tr}} \rangle$ & $D_{\diamond}$ & $\langle 1 - F \rangle / 10^3$ & $\langle D_{\textrm{tr}} \rangle /10^3 $ & $D_{\diamond} /10^3$ \\ \hline
Pol$_{\pi/8}$C & $ \left ( 0.67   \pm 0.30  \right ) p$ & $ (0.79 \pm 0.23) p$ & $p$ & $(2.26 \pm 0.81)  p^2$ & $(2.51 \pm 0.69) p^2$ & $3.39p^2$ \\ \hline
PCa & $  \left ( 0.67  \pm 0.24   \right )  p$ & $(0.74 \pm 0.20) p$ & $p$ & $ (2.26 \pm 0.81)  p^2$ & $(2.51 \pm 0.69) p^2$ & $3.39p^2$ \\ \hline
PCw & $  \left ( 0.90  \pm 0.19  \right )  p$ & $(0.93 \pm 0.18) p$ & $1.35p$ & $ (2.87 \pm 0.83)  p^2$ & $(3.06 \pm 0.76) p^2$ & $4.30p^2$ \\ \hline
CMCa & $  \left ( 0.71 \pm 0.25  \right )  p$ & $(0.78 \pm 0.22) p$ & $1.06p$ & $ (2.76 \pm 0.91)  p^2$ & $(3.02 \pm 0.79) p^2$ & $4.14p^2$ \\ \hline
CMCw & $  \left ( 0.78  \pm 0.26  \right )  p$ & $(0.86 \pm 0.23) p$ & $1.17p$ & $ (3.2 \pm 1.1)  p^2$ & $(3.53 \pm 0.92) p^2$ & $4.86p^2$ \\ \hline
DC & $p / 3$ & $2p/3$ & $p$ & $ (1.94 \pm 0.37) p^2$ & $(1.99 \pm 0.38) p^2$ & $2.91p^2$ \\ \hline
\end{tabular}
\end{center} \label{table:error_magnitude_PolXY}
\end{table*}

Table \ref{table:process_matrix_PolXY} presents the process matrices for the Pol$_{\pi/8}$C and its approximations at the physical level and the logical level with faulty EC.  Table \ref{table:error_magnitude_PolXY} describes how the error magnitude scales with $p$, the depolarizing strength.  

The trends are very similar to the ones observed on the ADC.  In this case, at the physical level, all the non-zero entries in the process matrices have linear leading orders, which means once again that the error magnitude is linear in the error strength.  In terms of the average trace distance, at the physical level, the constrained approximations are honest, while the unconstrained ones are dishonest.  However, this does not hold for the average error rate.  First, as expected, the average error rate of the Pol$_{\pi/8}$C is equal to the PCa's.  Also, the CMCa results in an honest approximation in terms of the average error rate.  

As observed for the ADC and its approximations, at the logical level, the leading terms in the process matrices become quadratic for the diagonal entries, but cubic for the off-diagonal ones.  This implies that, just like for the ADC, the advantage of the CMC approximations at the physical level becomes a drawback at the logical level, since the diagonal terms are more important than the off-diagonal.

There is an interesting difference between the effective process matrices at the logical level for the ADC and the Pol$_{\pi/8}$C.  For the former, the zero entries at the physical level remained zero at the logical level.  For the latter, however, this is not true.  For example, whereas the $\chi_{zz}$ entry is zero at the physical level for the Pol$_{\pi/8}$C and some of its approximations, at the logical level there is a term proportional to $p^2$.  The PCa provides the most intuitive case to understand where this term is coming from.  At the physical level, only $X$ and $Y$ errors occur.  However, for the Steane code with perfect EC, certain combinations of 1 $X$ and 2 $Y$ errors can result in an uncorrectable logical $Z$ error (For example, $IIIXYYI$), which give rise to a term in the $\chi_{zz}$ entry proportional to the third power of the error strength. On the other hand, if the EC is faulty, a $Y$ error on the ancillae and another one on the data can cause a logical $Z$ error, thus turning the error strength proportional to $p^2$.           

To summarize, we have found several common features for the 2 incoherent channels analyzed:
\begin{enumerate}
\item The expanded (CMC) channels provide more accurate approximations than the Pauli channels at the physical level.  However, at the logical level, they become completely eclipsed by the high accuracy of the PCa.  This characteristic of the PCa has been observed previously \cite{Geller_and_Zhou, Cory2014, PRA_us2015}.
\item The high accuracy of the PCa arises, somewhat ironically, from its inability to approximate the off-diagonal terms of the target error process matrix.  At the logical EC level, the off-diagonal terms in the process matrix are weaker (proportional to the cube of the error strength) than the diagonal terms (proportional to the square of the error strength), so in the low noise limit, the important contribution to the error at the logical level  is really made by the diagonal terms.   
\item At the physical level, the error magnitude for all the channels is linear in the error strength, whereas at the logical level, the error magnitude is quadratic in the error strength.  This holds for every error metric we have studied.  The average trace distance is consistently about twice as the average error rate, while the diamond distance is 3-4 times larger than the average error rate.
\end{enumerate}

\subsection{Coherent channels}

We define a coherent channel as a quantum operation that maps pure states to pure states.  They correspond to unitary rotations about a given axis in the Hilbert space of the system.  As model coherent errors, we have selected rotations about the Bloch sphere's $Z$ and $H$ (Hadamard) axes by an angle $\theta$:

\begin{equation}
\textrm{RZC} = \exp(-i\theta Z/2) = \cos(\theta/2) I  - i\sin(\theta/2) Z
\end{equation}
\begin{equation}
\textrm{RHC} = \exp(-i\theta H/2) = \cos(\theta/2) I  - i\sin(\theta/2) H
\end{equation}

These channels arise if there is an unwanted Hamiltonian during the gates, for example, from an uncontrolled magnetic field.  The angle $\theta$ parametrizes the error strength.  Just like for the incoherent channels, we have selected 150 uniformly distributed states on the Bloch sphere surface to calculate the average error rate and the average trace distance.

\subsubsection{RZC}

\begin{table*}[htdp]
\caption[RZC and approximations]
  {Process matrices for the RZC and its approximations at physical level and logical level with EC in the limit of small rotation angle ($\theta \rightarrow 0$).  Only the leading orders are shown.}
\begin{center}
\begin{tabular}{ c c c }
\hline \hline
\multicolumn{1}{ c }{Channel} & \multicolumn{1}{ c }{Physical process matrix} & \multicolumn{1}{ c }{Effective process matrix at the logical level} \\ \hline
\\
RZC & $\setlength{\arraycolsep}{14.5pt} \begin{pmatrix} 2 - \mathcal{O}(\theta^2)  & 0 & 0 & i\theta \\ 0 & 0 & 0 & 0 \\ 0 & 0 & 0 & 0 \\ -i\theta & 0 & 0 & \theta^2/2 \end{pmatrix}$ &  $\setlength{\arraycolsep}{16.75pt} \begin{pmatrix}  2 - \mathcal{O}(\theta^4) & 0 & 0 & i558 \, \theta^3 \\ 0 & 0 & 0 & 0 \\ 0 & 0 & 0 & 0 \\ -i558 \thinspace \theta^3 & 0 & 0 & 7870 \thinspace \theta^4 \end{pmatrix}$  \\
\\
PCa & $\setlength{\arraycolsep}{14.5pt} \begin{pmatrix} 2 - \mathcal{O}(\theta^2) & 0 & 0 & 0 \\ 0 & 0 & 0 & 0 \\ 0 & 0 & 0 & 0 \\ 0 & 0 & 0 & \theta^2/2 \end{pmatrix}$ &  $\setlength{\arraycolsep}{17.5pt} \begin{pmatrix} 2 - \mathcal{O}(\theta^4)  & 0 & 0 & 0 \\  0& 0 & 0 & 0 \\ 0 & 0 & 0 & 0 \\ 0 & 0 & 0 & 206 \thinspace \theta^4 \end{pmatrix}$  \\
\\
PCw & $\setlength{\arraycolsep}{17pt} \begin{pmatrix} 2 - \mathcal{O}(\theta) & 0 & 0 & 0 \\ 0 & 0 & 0 & 0 \\ 0 & 0 & 0 & 0 \\ 0 & 0 & 0 & \theta \thickspace \end{pmatrix}$ &   $\setlength{\arraycolsep}{17.3pt} \begin{pmatrix} 2 - \mathcal{O}(\theta^2)  &  0& 0 & 0 \\ 0 & 0 & 0 & 0 \\ 0 & 0 & 0 & 0 \\ 0 & 0 & 0 & 824 \thinspace \theta^2 \end{pmatrix}$ \\
\\
CMCa & $\setlength{\arraycolsep}{15pt} \begin{pmatrix}  2 - \mathcal{O}(\theta) & 0 & 0 & i\theta/2 \\ 0 & 0 & 0 & 0 \\ 0 & 0 & 0 & 0 \\ -i\theta/2 & 0 & 0 & \theta/2 \end{pmatrix}$ &   $\setlength{\arraycolsep}{16.7pt} \begin{pmatrix}  2 - \mathcal{O}(\theta^2) & 0 & 0 & i69.7 \thinspace \theta^3 \\ 0 & 0 & 0 & 0 \\ 0 & 0 & 0 & 0 \\ -i69.7 \thinspace \theta^3 &  0& 0 & 206 \thinspace \theta^2 \end{pmatrix}$  \\
\\
CMCw & $\setlength{\arraycolsep}{13.7pt} \begin{pmatrix}  2 - \mathcal{O}(\theta) & 0 & 0 & i\theta/\sqrt{2} \\ 0 & 0 & 0 & 0 \\ 0 & 0 & 0 & 0 \\ -i\theta/\sqrt{2} & 0 & 0 & \theta/\sqrt{2} \end{pmatrix}$ &   $\setlength{\arraycolsep}{16.7pt} \begin{pmatrix}  2 - \mathcal{O}(\theta^2) & 0 & 0 & i197 \thinspace \theta^3 \\ 0 & 0 & 0 & 0 \\ 0 & 0 & 0 & 0 \\ -i197 \thinspace \theta^3 & 0 & 0 & 412 \thinspace \theta^2 \end{pmatrix}$ \\
\\
DC & $\setlength{\arraycolsep}{9pt} \begin{pmatrix} 2 - \mathcal{O}(\theta^2) &0 & 0 & 0 \\ 0 & \theta^2/6 & 0 & 0 \\ 0 & 0 & \theta^2/6 & 0 \\ 0 & 0 & 0 & \theta^2/6 \end{pmatrix}$ &   $\setlength{\arraycolsep}{9pt} \begin{pmatrix} 2 - \mathcal{O}(\theta^4) & 0 & 0 & 0 \\ 0 & 193 \, \theta^4 & 0 & 0 \\ 0 & 0 & 20.1 \, \theta^4 & 0 \\ 0 & 0 & 0 & 150 \, \theta^4 \end{pmatrix}$ 
\\ \\ \hline \hline
\end{tabular}
\end{center} \label{table:process_matrix_RZC}
\end{table*}

\begin{table*}[htdp]
\caption[Behavior of the RZC and its approximations at various levels]
{Behavior of the RZC and its approximations at various levels in limit of small rotation angle ($\theta \rightarrow 0$) for the 3 different  error metrics. Only the leading orders are shown.  Standard deviations smaller than $10^{-7}$ are not presented.}
\begin{center}
\begin{tabular}{| c || c | c | c | c | c | c |}
\hline
\multicolumn{1}{| c ||}{Channel} & \multicolumn{3}{| c |}{Physical level} & \multicolumn{3}{| c |}{Logical level} \\ \cline{2-7}
 & $\langle 1 - F \rangle $ & $\langle D_{\textrm{tr}} \rangle $ & $D_{\diamond}$ & $\langle 1 - F \rangle / 10^3$ & $\langle D_{\textrm{tr}} \rangle / 10^3$ & $D_{\diamond} /10^3$ \\ \hline
RZC & $ \left ( 0.167 \pm 0.075 \right )  \thinspace \theta^2$ & $ \left ( 0.39  \pm 0.11 \right )   \thinspace \theta$ & $\theta / 2$ & $ (2.6 \pm 1.2) \thinspace  \theta^4$ & $(0.214 \pm 0.063) \thinspace \theta^3$ & $0.273 \thinspace  \theta^3$  \\ \hline
PCa & $ \left ( 0.167 \pm 0.075 \right )  \thinspace  \theta^2$ & $ \left ( 0.196  \pm 0.056 \right )  \thinspace \theta^2$ & $\theta^2 / 4$ & $(0.069 \pm 0.031) \thinspace  \theta^4$ & $(0.081 \pm 0.024) \thinspace \theta^4$ & $0.103 \thinspace \theta^4$ \\ \hline
PCw & $ \left ( 0.33  \pm 0.15 \right )  \thinspace \theta$ & $ \left ( 0.39  \pm 0.11 \right )  \thinspace \theta$ & $\theta / 2$ & $(0.28 \pm 0.13) \thinspace  \theta^2$ & $(0.328 \pm 0.097) \thinspace \theta^2$ & $0.420 \thinspace \theta^2$ \\ \hline
CMCa & $ \left ( 0.167 \pm 0.075 \right )  \thinspace \theta$ & $  \left ( 0.278  \pm 0.079 \right ) \thinspace \theta$ & $0.354 \thinspace \theta$ & $(0.071 \pm 0.032) \thinspace  \theta^2$ & $(0.083 \pm 0.025) \thinspace \theta^2$ & $0.107 \thinspace \theta^2$ \\ \hline
CMCw & $ \left ( 0.24 \pm 0.11 \right )  \thinspace \theta$ & $ \left ( 0.39  \pm 0.11 \right )  \thinspace \theta$ & $ \theta / 2$ & $(0.143 \pm 0.064) \thinspace  \theta^2$ & $(0.168 \pm 0.049) \thinspace \theta^2$ & $0.214 \thinspace \theta^2$ \\ \hline
DC & $\theta^2 / 6$ & $\theta^2 / 6$  & $\theta^2 / 4$ & $ (0.121 \pm 0.023) \thinspace  \theta^4$ & $ \left (0.125 \pm 0.024 \right )  \thinspace \theta^4$ & $0.182 \thinspace \theta^4$ \\ \hline
\end{tabular}
\end{center} \label{table:error_magnitude_RZC}
\end{table*}

Table \ref{table:process_matrix_RZC} presents the process matrices for the RZC and its approximations at the physical level and logical level with faulty EC.  Table \ref{table:error_magnitude_RZC} describes how the error magnitude scales with $\theta$, the over-rotation angle, in the limit $\theta \rightarrow 0$.  

The process matrix for the RZC reveals very different characteristics from those seen in the incoherent channels.  At the physical level, the process matrix has a quadratic term on its diagonal and a linear one on its off-diagonal entries.  Since the average error rate depends only on the diagonal terms, this implies, as seen in Table \ref{table:error_magnitude_RZC}, that there is a considerable mismatch between the average error rate and the distance-based metrics.  While for both the average trace distance and the diamond distance the error magnitude is linear in $\theta$, for the average error rate it is quadratic in $\theta$.  This agrees with the scaling of the diamond distance with the average error rate reported by Kueng \textit{et al.} \cite{Flammia_diamondvsfidelity2015} and Wallman \cite{Wallman_boundingerrors2014}.  

At the logical level, although not completely coherent anymore, the effective process matrix for the RZC still holds some features of its unitary nature at the physical level.  In this case, the diagonal term becomes quartic, while the off-diagonal ones becomes cubic in $\theta$.  This implies that the error magnitude becomes proportional to $\theta^4$ when quantified by the average error rate, but proportional to $\theta^3$ when quantified by a distance-based measure.               

These unique characteristics of the process matrices of the RZC make the approximation by the PCa problematic.  As expected, at the physical level the PCa matches the diagonal terms and the average error rate exactly, but this means it predicts a quadratic error, when in reality it is linear.  In terms of the average trace distance and the diamond distance, it underestimates the real error by one order of magnitude, making it very dishonest.  At the logical level, this extreme dishonesty is maintained.  Once again, in terms of the distance-based measures, the real error is underestimated by one order of magnitude ($\theta^4$ vs. $\theta^3$).  For the average error rate, something interesting occurs: although the error order is the same ($\theta^4$), because the off-diagonal terms in the physical process matrix influence the diagonal terms in the effective logical process matrix, the error magnitude is still severely underestimated.  As can be seen from Table \ref{table:error_magnitude_RZC}, at the logical level, the average error rate for the the RZC is about 37 times larger than for the PCa.  This highlights a very important limitation of the PCa: its inability to match off-diagonal terms turns it into a very bad approximate channel for coherent operations.

In order to not underestimate the error magnitude, at the physical level the PCw approximation results in a process matrix with a linear term on the diagonal.  This implies that the error magnitude is linear for every metric.  In terms of the average error rate, the error is overestimated by one order of magnitude ($\theta$ vs. $\theta^2$).  Interestingly, in terms of both the average trace distance and the diamond distance, the honesty constraint is tight: the PCw results in exactly the same error magnitude as the RZC.  At the logical level, however, the PCw becomes too pessimistic.  At this level, the error magnitude is proportional to $\theta^2$ for every error metric.

The expanded (CMC) approximations both result in process matrices at the physical level with linear terms on the off-diagonal and the diagonal entries.  Consequently, the error magnitude is linear for every error metric.  As expected, the CMCa approximation is dishonest when quantified with the average trace distance and the diamond distance.  However, it is honest by one order of magnitude when quantified with the average error rate.  The CMCw approximation is honest for every error metric, and just like the PCw, it saturates the honesty bound when quantified by the average trace distance and the diamond distance.  At the logical level, both approximate channels result in effective process matrices with a quadratic term on the diagonal and cubic terms on the off-diagonal entries, just like for the incoherent channels studied in the previous section.  This means that the error magnitude is proportional to $\theta^2$, regardless of the metric employed.  Both CMC approximations overestimate the error magnitude by 2 orders of $\theta$ when employing the average error rate, and 1 order of $\theta$ when employing the distance-based metrics.

\subsubsection{RHC}

\begin{table*}[htdp]
\caption[RHC and approximations]
  {Process matrices for the RHC and its approximations at physical level and logical level with EC in the limit of small rotation angle ($\theta \rightarrow 0$).  Only the leading orders are shown.}
\begin{center}
\begin{tabular}{ c c c }
\hline \hline
\multicolumn{1}{ c }{Channel} & \multicolumn{1}{ c }{Physical process matrix} & \multicolumn{1}{ c }{Effective process matrix at the logical level} \\ \hline
\\
RHC & $\setlength{\arraycolsep}{16pt} \begin{pmatrix} 2 - \mathcal{O}(\theta^2)  & i\theta / \sqrt{2} & 0 & i\theta/ \sqrt{2} \\ -i\theta / \sqrt{2} & \theta^2 / 4 & 0 & \theta^2 /4 \\ 0 & 0 & 0 & 0 \\ -i\theta /\sqrt{2} & \theta^2 / 4 & 0 & \theta^2/4 \end{pmatrix}$ &  $\setlength{\arraycolsep}{7.5pt} \begin{pmatrix}  2 - \mathcal{O}(\theta^4) & i828 \thinspace \theta^3  & \mathcal{O}(\theta^6) & i819 \thinspace \theta^3 \\ -i828 \thinspace \theta^3 &(1.27 \times 10^4) \thinspace \theta^4 & \mathcal{O}(\theta^7) & \mathcal{O}(\theta^6) \\  \mathcal{O}(\theta^6) & \mathcal{O}(\theta^7)  & \mathcal{O}(\theta^8) & \mathcal{O}(\theta^7) \\  -i819 \thinspace \theta^3 & \mathcal{O}(\theta^6) &\mathcal{O}(\theta^7)  & (1.22 \times 10^4) \thinspace \theta^4 \end{pmatrix}$  \\
\\
PCa & $\setlength{\arraycolsep}{18.25pt} \begin{pmatrix} 2 - \mathcal{O}(\theta^2) & 0 & 0 & 0 \\ 0 & \theta^2/4 & 0 & 0 \\ 0 & 0 & 0 & 0 \\ 0 & 0 & 0 & \theta^2/4 \end{pmatrix}$ &  $\setlength{\arraycolsep}{18.25pt} \begin{pmatrix} 2 - \mathcal{O}(\theta^4) & 0 & 0 & 0 \\ 0 & 107 \thinspace \theta^4 & 0 & 0 \\ 0 & 0 & \mathcal{O}(\theta^8) & 0 \\ 0 & 0 & 0 & 83.0 \thinspace \theta^4 \end{pmatrix}$  \\
\\
PCw & $\setlength{\arraycolsep}{18.75pt} \begin{pmatrix} 2 - \mathcal{O}(\theta) & 0 & 0 &0  \\0  &\theta/2 &0  &0  \\0  &0  & \theta/2 &0  \\ 0 & 0 & 0 & \theta/2 \thickspace \end{pmatrix}$ &   $\setlength{\arraycolsep}{17pt} \begin{pmatrix} 2 - \mathcal{O}(\theta^2)  & 0 &0  & 0 \\ 0 & 1710 \thinspace \theta^2 & 0 & 0 \\ 0 & 0 & 186 \thinspace \theta^2  & 0 \\ 0 & 0 & 0 & 1340 \thinspace \theta^2 \end{pmatrix}$ \\
\\
CMCa & $\setlength{\arraycolsep}{14.75pt} \begin{pmatrix}  2 - \mathcal{O}(\theta) & i0.283 \thinspace \theta & 0 & i0.283 \thinspace \theta \\ -i0.283 \thinspace \theta & 0.283 \thinspace \theta & 0 & 0 \\ 0 & 0 & 0 & 0 \\ -i0.283 \thinspace \theta & 0 & 0 & 0.283 \thinspace \theta \end{pmatrix}$ &   $\setlength{\arraycolsep}{11pt} \begin{pmatrix}  2 - \mathcal{O}(\theta^2) & i52.1 \thinspace \theta^3 & \mathcal{O}(\theta^5) & i52.1 \thinspace \theta^3 \\  -i52.1 \thinspace \theta^3 &136 \thinspace \theta^2 & \mathcal{O}(\theta^6) & \mathcal{O}(\theta^5) \\  \mathcal{O}(\theta^5) &  \mathcal{O}(\theta^6) & (1.03 \times 10^4) \thinspace \theta^4 & \mathcal{O}(\theta^6) \\  -i52.1 \thinspace \theta^3 & \mathcal{O}(\theta^5) & \mathcal{O}(\theta^6) & 106 \thinspace \theta^2 \end{pmatrix}$  \\
\\
CMCw & $\setlength{\arraycolsep}{11.25pt} \begin{pmatrix}  2 - \mathcal{O}(\theta) & i0.408 \, \theta   & 0  & i0.408 \, \theta   \\ -i0.408 \, \theta  & 0.408 \, \theta & 0  & 0.204 \, \theta  \\ 0 & 0  & 0.204 \, \theta  & 0 \\ -i0.408 \, \theta  & 0.204 \, \theta & 0  & 0.408 \, \theta  \end{pmatrix}$ &   $\setlength{\arraycolsep}{13.5pt} \begin{pmatrix}  2 - \mathcal{O}(\theta^2) & i159 \, \theta^3 & \mathcal{O}(\theta^4) & i155 \, \theta^3 \\ -i159 \, \theta^3 & 662 \, \theta^2 & \mathcal{O}(\theta^4) & -0.571 \, \theta^3 \\ \mathcal{O}(\theta^4) & \mathcal{O}(\theta^4) & 31.3 \, \theta^2 & \mathcal{O}(\theta^4) \\ -i155 \, \theta^3  & -0.571 \, \theta ^3 & \mathcal{O}(\theta^4) & 516 \, \theta^2  \end{pmatrix}$ \\
\\
DC & $\setlength{\arraycolsep}{16.5pt} \begin{pmatrix} 2 - \mathcal{O}(\theta^2) & 0 & 0 & 0 \\ 0 & \theta^2/6 & 0 & 0 \\ 0 & 0 & \theta^2/6 & 0 \\ 0 & 0 & 0 & \theta^2/6 \end{pmatrix}$ &   $\setlength{\arraycolsep}{18.5pt} \begin{pmatrix} 2 - \mathcal{O}(\theta^4) & 0 & 0 & 0 \\ 0 & 193 \thinspace \theta^4 & 0 & 0 \\ 0 & 0 & 20.1 \thinspace \theta^4 & 0 \\ 0 &0  &0  & 150 \thinspace \theta^4 \end{pmatrix}$ 
\\ \\ \hline \hline
\end{tabular}
\end{center} \label{table:process_matrix_RHC}
\end{table*}

\begin{table*}[htdp]
\caption[Behavior of the RHC and its approximations at various levels]
{Behavior of the RHC and its approximations at various levels in limit of small rotation angle ($\theta \rightarrow 0$) for the 3 different metrics.  Only the leading orders are shown.  Standard deviations smaller than $10^{-7}$ are not presented.}
\begin{center}
\begin{tabular}{| c || c | c | c | c | c | c |}
\hline
\multicolumn{1}{| c ||}{Channel} & \multicolumn{3}{| c |}{Physical level} & \multicolumn{3}{| c |}{Logical level} \\ \cline{2-7}
 & $\langle 1 - F \rangle $ & $\langle D_{\textrm{tr}} \rangle$ & D$_{\diamond} $ & $\langle 1 - F \rangle / 10^3$ & $\langle D_{\textrm{tr}} \rangle /10^3$ & D$_{\diamond} /10^3 $ \\ \hline
RHC & $(0.167 \pm 0.075) \thinspace  \theta^2$ & $(0.39 \pm 0.11) \thinspace  \theta$ & $\theta / 2$ & $(8.5 \pm 1.9) \thinspace  \theta^4$ & $(0.45 \pm 0.14) \thinspace  \theta^3$ & $0.589 \thinspace \theta^3$ \\ \hline
PCa & $(0.167 \pm 0.037) \thinspace  \theta^2$ & $(0.172 \pm 0.039) \thinspace \theta^2$ & $ \theta^2 / 4$ & $(0.063 \pm 0.015) \thinspace  \theta^4$ & $(0.066 \pm 0.015) \thinspace \theta^4$ & $0.0951 \thinspace \theta^4$ \\ \hline
PCw & $\theta / 2$ & $\theta / 2$ & $3 \, \theta /4$ & $(0.98 \pm 0.16) \thinspace  \theta^2$ & $(1.00 \pm 0.17) \thinspace \theta^2$ & $1.64 \thinspace  \theta^2$ \\ \hline
CMCa & $(0.189 \pm 0.042) \thinspace \theta$ & $(0.251 \pm 0.058) \thinspace \theta$ & $0.362 \thinspace \theta$ & $(0.084 \pm 0.020) \thinspace \theta^2$ & $(0.087 \pm 0.020) \thinspace \theta^2$ & $0.126 \thinspace \theta^2$ \\ \hline
CMCw & $(0.340 \pm 0.060) \, \theta$ & $(0.416 \pm 0.081) \, \theta $ & $ 0.602 \, \theta$ & $(0.397 \pm 0.080) \, \theta^2$ & $(0.408 \pm 0.081) \, \theta^2$ & $0.596 \, \theta^2$ \\ \hline
DC & $\theta^{2} / 6$ & $\theta^2 / 6$  & $\theta^2 /4$ & $ (0.121 \pm 0.023) \thinspace  \theta^4$ & $(0.125 \pm 0.024) \thinspace \theta^4$ & $0.182 \thinspace  \theta^4$ \\ \hline
\end{tabular}
\end{center} \label{table:error_magnitude_RHC}
\end{table*}

Table \ref{table:process_matrix_RHC} presents the process matrices for the RHC and its approximations at the physical level and logical level with faulty EC.  Table \ref{table:error_magnitude_RHC} describes how the error magnitude scales with $\theta$ in the limit of $\theta \rightarrow 0$.

The results are very similar to the RZC.  At the physical level, the process matrix has off-diagonal quadratic terms and linear diagonal terms.  This implies, once again, that the average error rate is quadratic in $\theta$, while the average trace distance and the diamond distance are linear in $\theta$.  

In contrast to the RZC, at the logical level, the effective process matrix for the RHC is considerably less sparse than at the physical level.  However, the important trends are maintained.  The strongest terms are proportional to $\theta^4$ on the diagonal and to $\theta^3$ on the off-diagonal.  Once again, because of its lack of access to the off-diagonal entries, the PCa provides an extremely optimistic (and dishonest) approximation both at the physical and logical levels.  

The PCw provides an honest approximation at the physical level, but this results in an overly pessimistic one at the logical level: it predicts the diamond distance to scale like $\theta^2$, when in reality it scales like $\theta^3$.  Regarding the expanded (CMC) approximations, at the logical level, they result in effective process matrices where the strongest terms scale like $\theta^2$ on the diagonal and $\theta^3$ on the off-diagonal.  This means that they overestimate the error magnitude by 2 orders of magnitude, when employing the average error rate, and 1 order of magnitude, when employing the distance-based metrics.  The expanded approximate channels (CMCs) show the same trends as in the RZC case.        

To summarize, the common features for the 2 coherent channels analyzed are:
\begin{enumerate}
\item Just like for the incoherent operations, the expanded (CMC) channels provide more accurate approximations then the Pauli channels at the physical level.
\item In contrast to incoherent operations, at the logical level the PCa results in a very bad approximation.  Formally, it is still the most accurate approximation, since its difference from the target channel is proportional to $\theta^3$ (the error magnitude of the target channel), while all the other approximations differ from the target channel by a term proportional to $\theta^2$ (the error magnitude of the approximate channel).  However, the PCa underestimates the error magnitude of the real error so severely that, for practical purposes, it is a very bad approximation.  This holds for every error metric used, but it is more pronounced for the average trace distance and the diamond distance.
\item The severe underestimation of the error magnitude by the PCa is caused by its lack of access to off-diagonal terms in the process matrix.  In contrast to incoherent channels, coherent rotations have process matrices whose off-diagonal terms are stronger by 1 order of the error strength, $\theta$, than the diagonal terms.  
\item In contrast to incoherent channels, no approximation results in a very accurate description at the logical level.  While the PCa severely underestimates the error magnitude of the target coherent channel, the rest of the approximations overestimate it, because of this particular trait of coherent channels of having stronger off-diagonal than diagonal terms.  At the logical level, we cannot approximate accurately the behavior of the coherent channels, at least not with the methods developed in \cite{PRA_us}.  However, we can use the PCa to provide a lower bound and the other approximate channels to provide an upper bound to the error magnitude.  The CMCa (unconstrained expanded approximation) provides the tighest upper bound.
\item In contrast to incoherent channels, the error magnitude of coherent channels greatly depends on the metric employed.  At the physical level, the average error rate returns a quadratic error, while the distance-based metrics return a linear error.  At the logical level, the average error rate returns a quartic error, while the distance-based metric return a cubic error.  This occurs because the average error rate depends exclusively on the diagonal terms of the process matrix, which are weaker than the off-diagonal terms for coherent channels.          
\item Comparing the orders of the terms at the diagonal and off-diagonal entries in the process matrix provides a convenient way to visualize how the diamond distance scales with the average error rate. At the physical level, the diamond distance scales like the square root of the average error rate ($D_{\diamond} \propto e^{1/2}$).  However, at the logical level, the exponent of the scaling becomes $3/4$ ($D_{\diamond} \propto e^{3/4}$).  As shown by Kueng \textit{et al.} \cite{Flammia_diamondvsfidelity2015} and Wallman \cite{Wallman_boundingerrors2014}, a scaling exponent of $1/2$ is a characteristic signature of coherent noise.  On the other hand, completely Pauli noise has a scaling exponent of $1$.  Therefore, at the logical level, the effective noise, although not yet Pauli, becomes less coherent than the physical noise.     
\end{enumerate}

\section{Level-1 pseudo-thresholds}
\label{sec:thresholds}

As in \cite{PRA_us2015}, the pseudo-threshold is calculated by finding the intersection between the error magnitude curve at the physical level with the error magnitude curve at the logical level.  For the approximate channels, we simulate the real error exactly at the physical level and use the approximations only at the logical level.  We follow the notation convention of \cite{PRA_us2015} and refer to these as \textit{exact channel } / \textit{approximate channel}.  This makes the comparisons between the different thresholds estimates easier, since they will only depend on the behavior of the approximate channels at the logical level.  We report the pseudo-thresholds obtained by the different error magnitude metrics.

\begin{table}[htdp]
\caption[Threshold ADC]
{Level-1 pseudo-thresholds for the ADC and its approximations for the 3 different error magnitude metrics.}
\begin{center}
\begin{tabular}{| c || c | c | c |}
\hline
\multicolumn{1}{| c ||}{Channel} & \multicolumn{1}{| c |}{ $\langle 1 - F \rangle $ } & \multicolumn{1}{| c |}{ $\langle D_{\textrm{tr}} \rangle$  } & \multicolumn{1}{| c |}{$D_{\diamond} $  } \\ \cline{2-4}
& $\gamma_{\textrm{th}} \times 10^{4}$ & $\gamma_{\textrm{th}} \times 10^{4}$ & $\gamma_{\textrm{th}} \times 10^{4}$ \\ \hline
ADC & $4.8 \pm 4.2$ & $7.7 \pm 4.3$ & $9.10$ \\ \hline
ADC / PCa & $4.9 \pm 4.2$ & $7.7 \pm 4.3$ & $9.24$ \\ \hline
ADC / PCw & $0.97 \pm 0.86$ & $1.56 \pm 0.84$ & $1.86$ \\ \hline
ADC / CMCa & $5.8 \pm 5.6$ & $9.2 \pm 5.1$ & $11.0$ \\ \hline
ADC / CMCw & $3.2 \pm 3.1$ & $5.2 \pm 2.8$ & $6.16$ \\ \hline
ADC / DC & $7.4 \pm 7.2$ & $11.8 \pm 6.5$  & $14.1$ \\ \hline
\end{tabular}
\end{center} \label{table:threshold_ADC}
\end{table}

\begin{figure}[h!]
\includegraphics[scale=0.55]{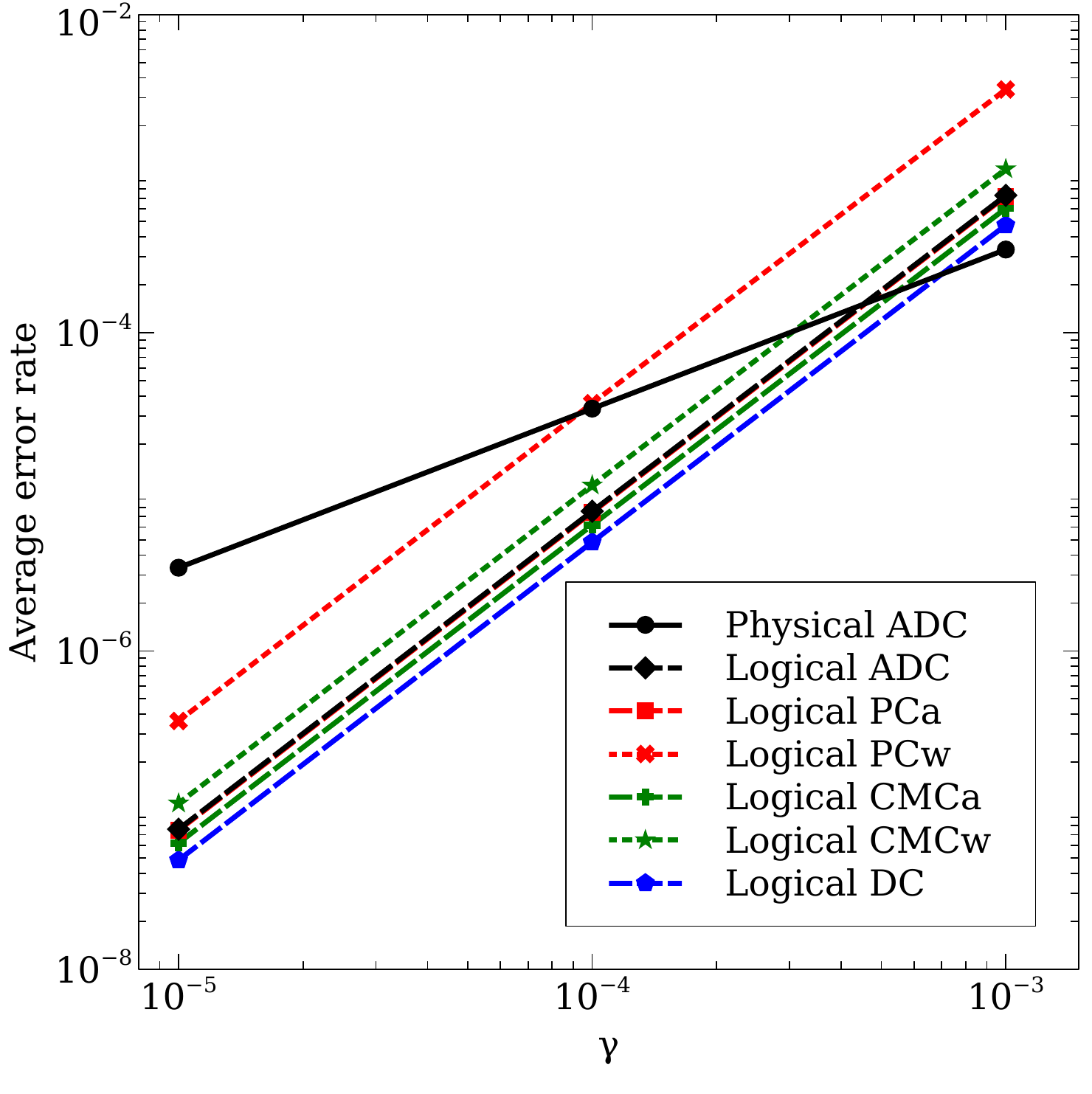}
\caption{Average error rate for the ADC and its approximations for various damping strengths.  The curve for the PCa at the logical level is located exactly underneath the curve for the ADC at the logical level.  The level-1 pseudo-thresholds are given by the intersection between the best fits for the physical and the logical error magnitudes.  The lines are just guides to the eye and do not correspond to the best fits.}
\label{fig:ADC_fidelity}
\end{figure}

\begin{figure}[h!]
\includegraphics[scale=0.55]{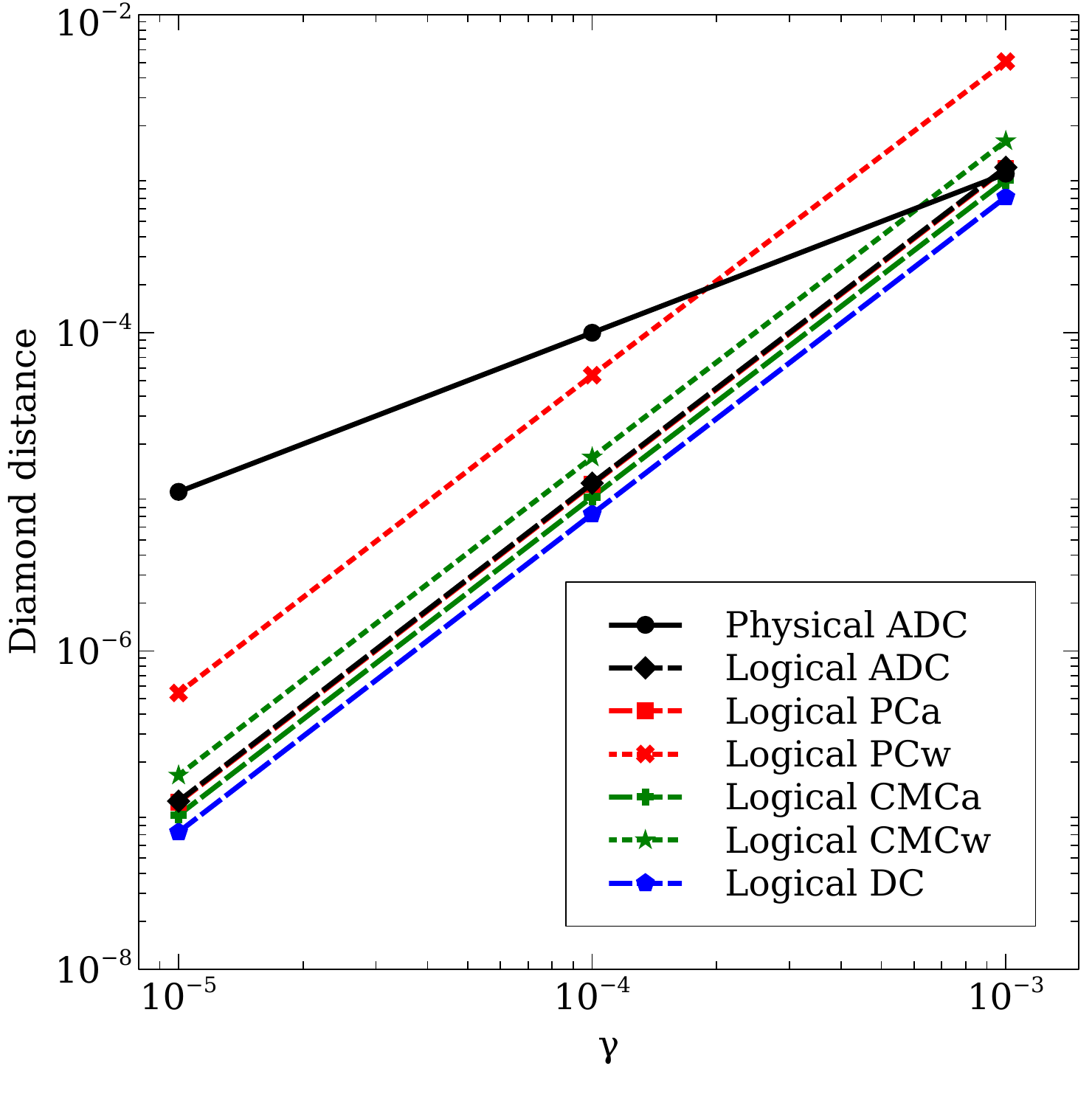}
\caption{Diamond distance for the ADC and its approximations for various damping strengths.  The curve for the PCa at the logical level is located exactly underneath the curve for the ADC at the logical level.  The lines are guides to the eye.}
\label{fig:ADC_diamond}
\end{figure}

\begin{table}[htdp]
\caption[Threshold PolXY]
{Level-1 pseudo-threshold Pol$_{\pi/8}$C and its approximations for the 3 different error magnitude metrics.}
\begin{center}
\begin{tabular}{| c || c | c | c |}
\hline
\multicolumn{1}{| c ||}{Channel} & \multicolumn{1}{| c |}{ $\langle 1 - F \rangle $ } & \multicolumn{1}{| c |}{ $\langle D_{\textrm{tr}} \rangle$  } & \multicolumn{1}{| c |}{$D_{\diamond} $  } \\ \cline{2-4}
& $p_{\textrm{th}} \times 10^{4}$ & $p_{\textrm{th}} \times 10^{4}$ & $p_{\textrm{th}} \times 10^{4}$ \\ \hline
Pol$_{\pi/8}$C & $3.1 \pm 1.2$ & $3.4 \pm 1.2$ & $3.02$ \\ \hline
Pol$_{\pi/8}$C / PCa & $3.1 \pm 1.2$ & $3.4 \pm 1.2$ & $3.02$ \\ \hline
Pol$_{\pi/8}$C / PCw & $2.32 \pm 0.87$ & $2.65 \pm 0.73$ & $2.36$ \\ \hline 
Pol$_{\pi/8}$C / CMCa & $2.45 \pm 0.91$ & $2.74 \pm 0.83$ & $2.46$ \\ \hline
Pol$_{\pi/8}$ / CMCw & $2.08 \pm 0.77$ & $2.33 \pm 0.69$ & $2.09$ \\ \hline
Pol$_{\pi/8}$C / DC & $3.50$ & $4.07$  & $3.49$ \\ \hline
\end{tabular}
\end{center} \label{table:threshold_PolXY}
\end{table}

Tables \ref{table:threshold_ADC} and \ref{table:threshold_PolXY} present the level-1 pseudo-thresholds for the ADC and for Pol$_{\pi/8}$C and their corresponding approximations, respectively.  The pseudo-thresholds are given in terms of the characteristic error strength: the damping strength ($\gamma$) for the ADC and the error probability $p$ for the Pol$_{\pi/8}$C.  The pseudo-threshold values in terms of the average error rate, the average trace distance, and the diamond distance can be easily calculated from the expressions on Tables \ref{table:error_magnitude_ADC} and \ref{table:error_magnitude_PolXY}.  Figure \ref{fig:ADC_fidelity} shows how the average error rate scales with the damping strength, $\gamma$, for the ADC and its approximations.  Figure \ref{fig:ADC_diamond} shows the scaling of the diamond distance.  The pseudo-threshold for the ADC and its approximations can be visually obtained from these.

There are important common characteristics for both incoherent channels.  First, and as observed in \cite{PRA_us2015}, the PCa estimates the pseudo-threshold very accurately.  Its pseudo-thresholds are practically the same as the exact ones regardless of the error metric used.  Second, since the constrained (``w'') channels remain honest at the logical level, they always provide lower bounds to the pseudo-threshold.  This is not that useful when approximating incoherent channels, since the PCa is so accurate, but it will be useful for the coherent errors.   Finally, the pseudo-threshold does not depend too much on the error metric.  In the case of the ADC, the diamond distance pseudo-threshold ($9.1 \times 10^{-4}$) is almost twice as the average error rate pseudo-threshold ($4.8 \times 10^{-4}$), but they are both in the same order of magnitude.  For the Pol$_{\pi/8}$C, the pseudo-thresholds obtained from the different metrics are practically the same.

\begin{table}[htdp]
\caption[Threshold RZC]
{Level-1 pseudo-thresholds for the RZC and its approximations for the 3 different error magnitude metrics.}
\begin{center}
\begin{tabular}{| c || c | c | c |}
\hline
\multicolumn{1}{| c ||}{Channel} & \multicolumn{1}{| c |}{ $\langle 1 - F \rangle $ } & \multicolumn{1}{| c |}{ $\langle D_{\textrm{tr}} \rangle$  } & \multicolumn{1}{| c |}{$D_{\diamond} $  } \\ \cline{2-4}
& $\theta_{\textrm{th}} \times 10^{3}$ & $\theta_{\textrm{th}} \times 10^{3}$ & $\theta_{\textrm{th}} \times 10^{3}$ \\ \hline
RZC & $7.92$ & $39.8$ & $39.8$ \\ \hline
RZC / PCa & $49.5$ & $170$ & $170$ \\ \hline
RZC / PCw & $0$ & $1.22$ & $1.22$ \\ \hline
RZC / CMCa & $0$ & $4.84$ & $4.84$ \\ \hline
RZC / CMCw & $0$ & $2.40$ & $2.40$ \\ \hline
RZC / DC & $36 \pm 11$ & $147 \pm 24$  & $142$ \\ \hline
\end{tabular}
\end{center} \label{table:threshold_RZC}
\end{table}

\begin{figure}[h!]
\includegraphics[scale=0.55]{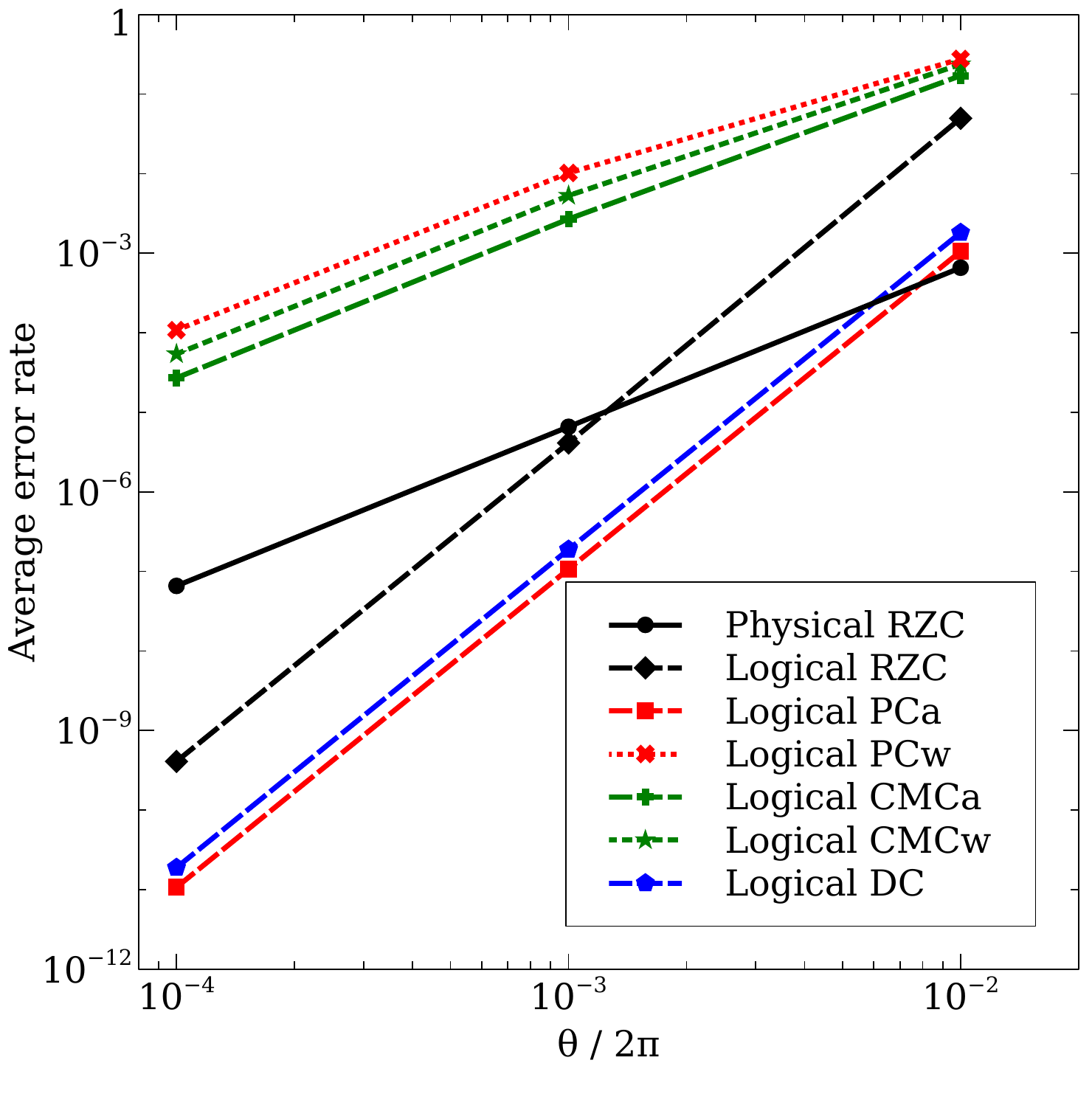}
\caption{Average error rate for the RZC and its approximations for various over-rotation angles.  The curves for the PCw, CMCa, and CMCw do not intersect the physical RZC curve because they all scale quadratically with $\theta$, but the coefficient of the RZC curve is the smallest.  In this case, the PCw and the expanded channels provide a useless lower bound ($0$) to the exact pseudo-threshold.  The lines are guides to the eye.}
\label{fig:RZC_fidelity}
\end{figure}

\begin{figure}[h!]
\includegraphics[scale=0.55]{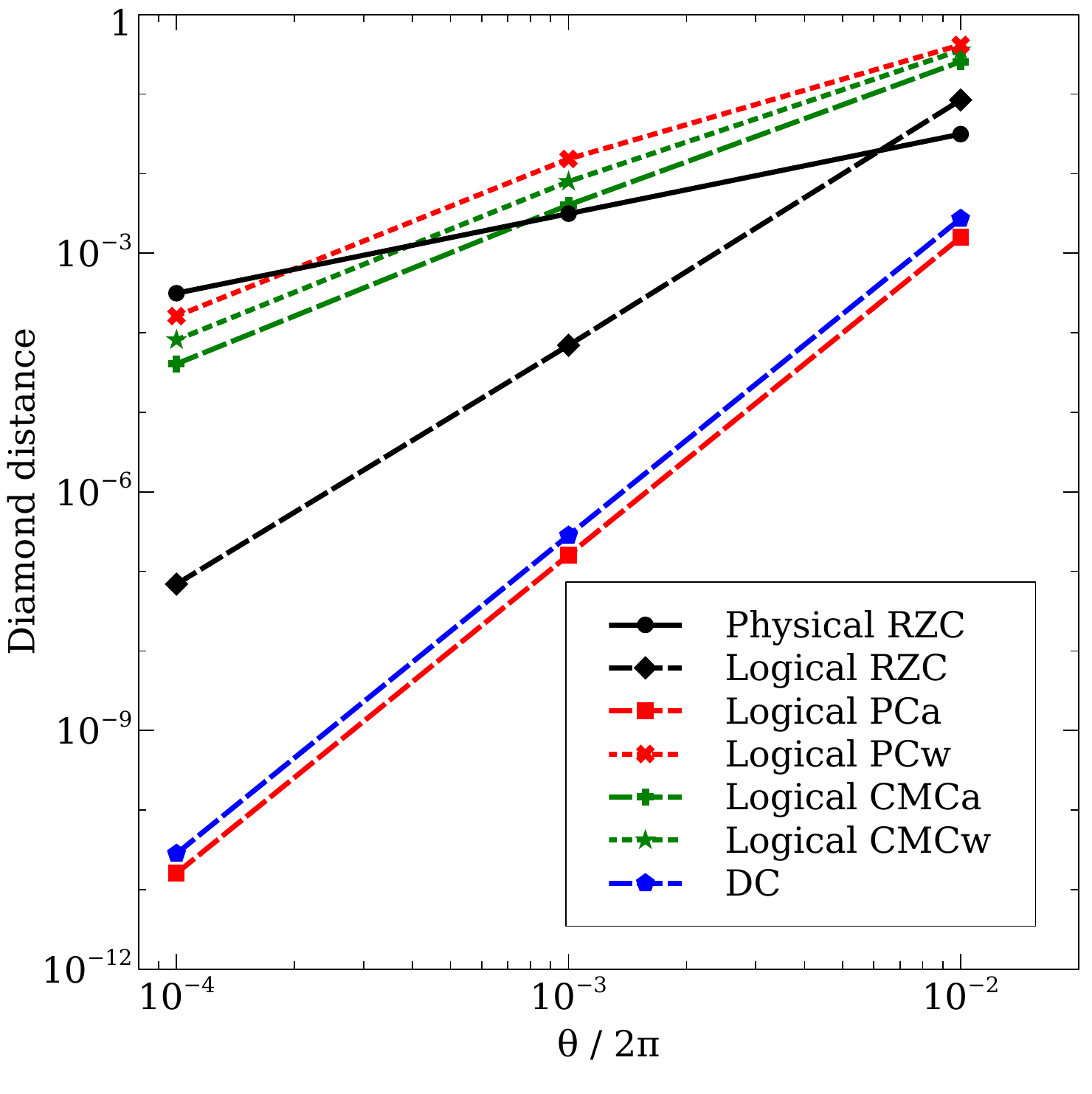}
\caption{Diamond distance for the RZC and its approximations for various over-rotation angles. The lines are guides to the eye.}
\label{fig:RZC_diamond}
\end{figure}

\begin{table}[htdp]
\caption[Threshold RHC]
{Level-1 pseudo-threshold for the RHC and its approximations for different error metrics.}
\begin{center}
\begin{tabular}{| c || c | c | c |}
\hline
\multicolumn{1}{| c ||}{Channel} & \multicolumn{1}{| c |}{ $\langle 1 - F \rangle $ } & \multicolumn{1}{| c |}{ $\langle D_{\textrm{tr}} \rangle$  } & \multicolumn{1}{| c |}{$D_{\diamond} $  } \\ \cline{2-4}
& $\theta_{\textrm{th}} \times 10^{3}$ & $\theta_{\textrm{th}} \times 10^{3}$ & $\theta_{\textrm{th}} \times 10^{3}$ \\ \hline
RHC & $4.3 \pm 1.2$ & $28.5 \pm 2.1$ & $27.7$ \\ \hline
RHC / PCa & $50 \pm 14$ & $181 \pm 21$ & $174$ \\ \hline
RHC / PCw & $0$ & $0.33 \pm 0.12$ & $0.272$ \\ \hline
RHC / CMCa & $0$ & $4.7 \pm 1.5$ & $4.07$ \\ \hline
RHC / CMCw & $0$ & $0.97 \pm 0.30$ & $0.837$ \\ \hline
RHC / DC & $36 \pm 11$ & $147 \pm 24$  & $142$ \\ \hline
\end{tabular}
\end{center} \label{table:threshold_RHC}
\end{table}

Tables \ref{table:threshold_RZC} and \ref{table:threshold_RHC} present the level-1 pseudo-threshold for the RZC and the RHC and their corresponding approximations, respectively.  The pseudo-thresholds are given in terms of the over-rotation angle, $\theta$.  Figure \ref{fig:RZC_fidelity} shows how the average error rate scales with the over-rotation angle for the RZC.  Figure \ref{fig:RZC_diamond} shows the scaling of the diamond distance. 

The level-1 pseudo-threshold trends are very different for coherent channels.  The PCa gives an overly optimistic estimate of the pseudo-threshold.  The PCw and the expanded channels give an overly pessimistic pseudo-threshold.  This holds for every error metric used.  In contrast to the incoherent case, none of our approximations predicts the pseudo-threshold accurately, although we can bound the pseudo-threhsold.  The PCa will always give an upper bound, while the PCw and expanded channels will always give a lower bound.  Notice, however, that this lower bound is not useful at all if we employ the average error rate as our metric.  In this case, the pseudo-threshold given by the PCw and the expanded approximations is exactly 0.  When the error is quantified by the average trace distance and the diamond distance, the lower bound is tighter.   

\begin{figure}[h!]
\includegraphics[scale=0.55]{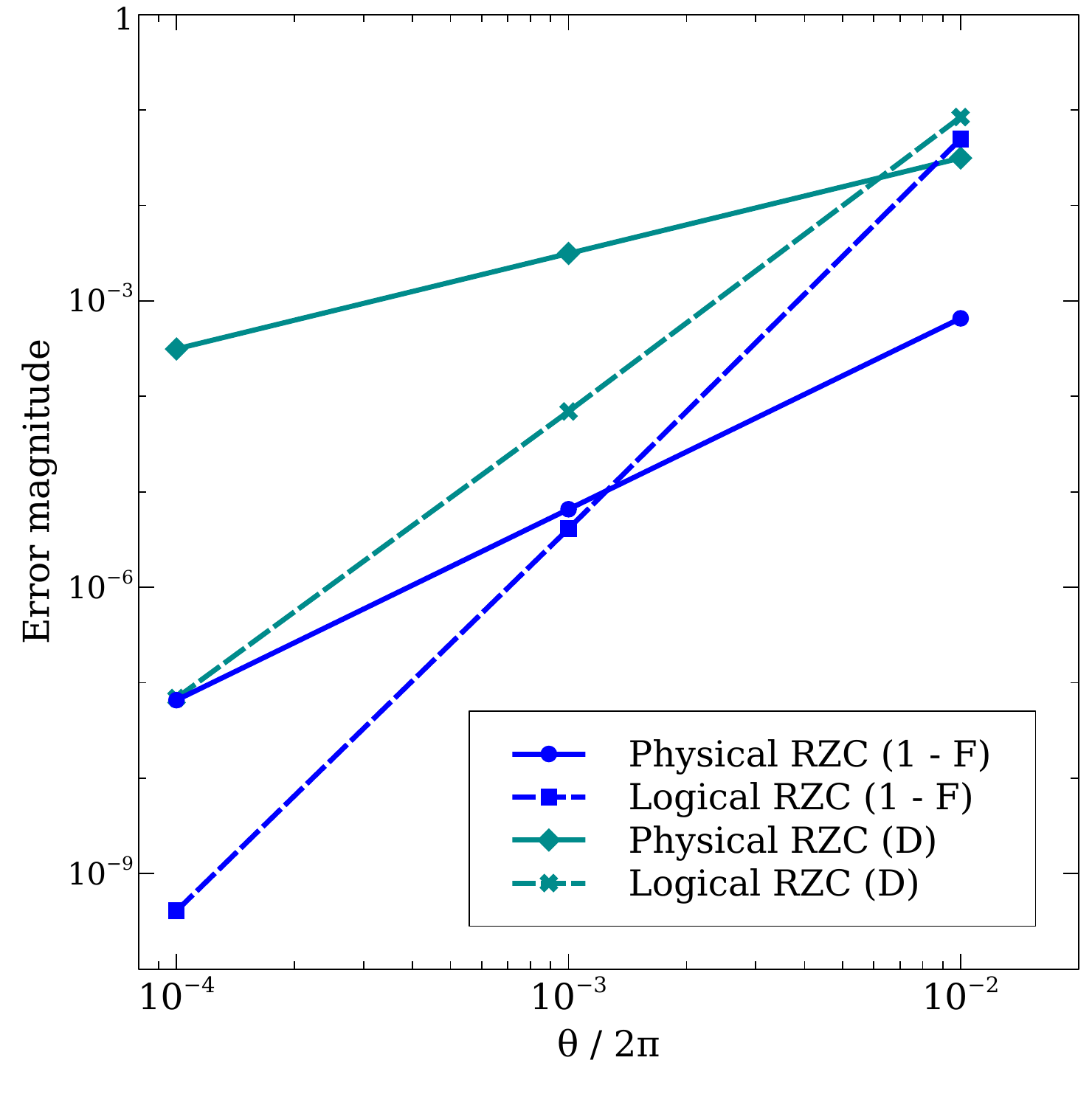}
\caption{Error magnitude for the RZC quantified with two different error metrics: the average error rate and the diamond distance.  The intersection of the two blue curves gives the level-1 pseudo-threshold based on the average error rate.  The intersection of the two cyan curves gives the pseudo-threshold based on the diamond distance.  In this case, the $\theta_{\textrm{th}}$ obtained from the diamond distance is about 1 order of magnitude higher than the one obtained from the average error rate, as can be seen on Table \ref{table:threshold_RZC}.}
\label{fig:RZC_real_f_D}
\end{figure}

\begin{figure}[h!]
\includegraphics[scale=0.55]{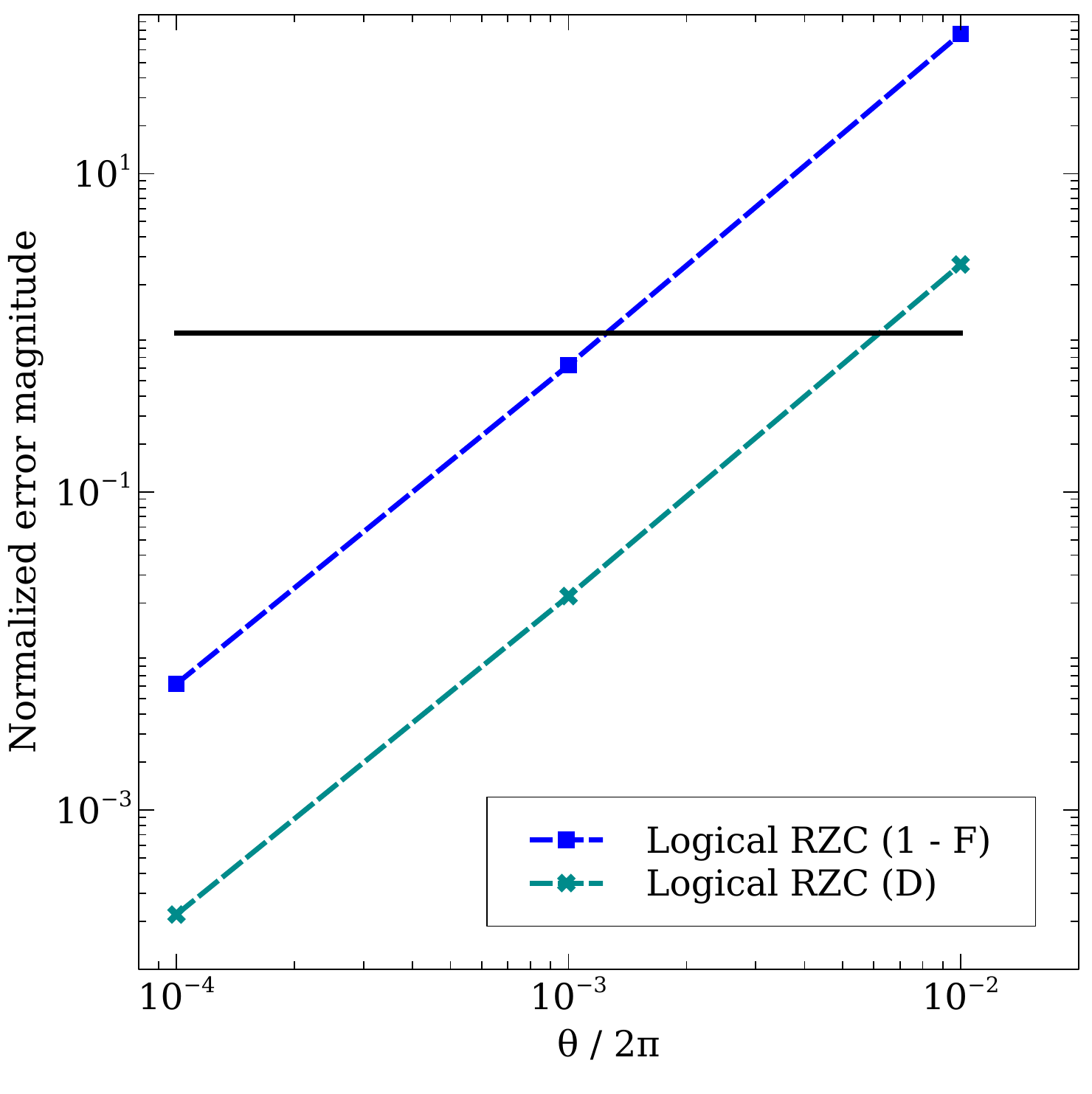}
\caption{Normalized error magnitude for the RZC.  The normalization factor is the respective error magnitude for each method at the physical level.}
\label{fig:RZC_f_D_norm}
\end{figure}

For coherent channels, the level-1 pseudo-threshold quantified by the diamond distance is 1 order of magnitude higher than the one quantified by the average error rate.  This is quite unexpected, since the assumption is that the threshold value given by the diamond distance would actually be lower than the one given by the average error rate, because the diamond distance is a worst-case measure.  Intuitively, this is a consequence of the average error rate being very small ($0.082 \, \theta^2$) and the diamond distance being considerably larger ($\theta/2$) at the physical level.  At the logical level, the diamond distance is still larger than the average error rate, but by a smaller amount. This can be observed on Figure \ref{fig:RZC_real_f_D}.  The appreciate how the different scalings cause a huge discrepancy between the two pseudo-thresholds, it is convenient to normalize the error magnitude at the logical level by the error magnitude at the physical level.  This is shown in Figure \ref{fig:RZC_f_D_norm}.  Since we are normalizing by the respective error magnitude at the physical level, both curves at this level become equal to 1 and the curves at the logical level become quadratic in $\theta$.

\section{Conclusions} \label{sec:conclusions} 

We have computed the physical process ($\chi$) matrix and the effective 1-qubit process ($\chi$) matrix at the first level of EC with the Steane [[7,1,3]] code for different incoherent and coherent error models in the limit of low noise.  For incoherent errors, at the logical level, the off-diagonal terms decay faster than the diagonal terms, which explains the high accuracy of the PCa for incoherent channels.  On the other hand, for coherent errors, the off-diagonal terms, stronger than the diagonal ones at the physical level, remain stronger at the logical level.  This implies that the PCa approximation (and any stochastic approximation that matches the average fidelity at the physical level) to coherent channels will unavoidably underestimate the error magnitude at the logical level.  On the other hand, a stochastic channel that matches (or does not underestimate) a distance-based measure at the physical level will result in an approximation that is too pessimistic at the logical level.  These trends provide bounds on the pseudo-threshold of coherent channels, but these are not very tight.  We have also observed that for coherent channels, the level-1 pseudo-threshold depends strongly on the error measure employed.  However, distance-based measures result in considerably higher pseudo-thresholds than the average fidelity.


As several authors have shown \cite{Wallman_unitarity, Flammia_diamondvsfidelity2015}, a characteristic feature of stochastic noise is that the diamond distance scales linearly with the average error rate.  In contrast, for coherent noise the diamond distance scales like the square root of the average error rate.  As we can see from the effective process matrices at the logical level, if the physical noise is unitary, at the first level of concatenation, the effective noise has $\theta^3$ off-diagonals terms and $\theta^4$ diagonal terms.  This means that diamond distance scales like $\langle 1 - F \rangle ^{3/4}$.  The scaling of this exponent as a function of the level of concatenation or the code distance remains an open question.

We have examined channels that are either coherent or incoherent while in reality most channels contain both aspects \cite{Wallman_unitarity}.  For the unitary error models considered, the errors can be completely removed by dynamic-decoupling techniques \cite{Viola1999, Khodjasteh2010, Biercuk2011}.  In general, open-loop control techniques can be used to transform error channels that have a larger coherent character into errors that have less coherent character \cite{Kabytayev2014, Soare2014}.  At the gate level the addition of Pauli twirling gates \cite{Emerson2007, Wallman_randcomp}  can also reduce the coherent noise. Given the negative effect of coherent errors on the pseudothreshold, we expect these coherent noise reducing methods will be essential for achieving logical qubits that outperform physical qubits.


\begin{acknowledgments}

We acknowledge discussions and contributions from Joel J. Wallman and Nathaniel Johnston.  This work was supported by the Office of the Director of National Intelligence - Intelligence Advanced Research Projects Activity through ARO contract W911NF-10-1-0231 and the National Science Foundation grant PHY-1415461.

\end{acknowledgments}

\newcommand{\noopsort}[1]{} \newcommand{\printfirst}[2]{#1}
  \newcommand{\singleletter}[1]{#1} \newcommand{\switchargs}[2]{#2#1}

\end{document}